\documentclass[a4paper,11pt]{article}
\pdfoutput=1 

\usepackage{jheppub} 

\usepackage[T1]{fontenc} 

\usepackage{graphicx}
\usepackage{epsfig}
\usepackage{color}
\usepackage{comment}
\usepackage{amsthm}
\usepackage{hyperref}
\usepackage{amssymb}
\usepackage{booktabs}
\usepackage{slashed}
\usepackage{physics}
\usepackage{microtype}
\usepackage{dsfont}

\title{Signals of Doomsday II: Cosmological signatures of late-time $SU(3)_c$ symmetry breaking}

\author[a,b]{Amartya Sengupta,}
\author[a]{Dejan Stojkovic,}
\author[b]{L.C.R. Wijewardhana}

\affiliation[a]{HEPCOS, Department\, of \,Physics, SUNY\, at\, Buffalo, Buffalo, NY\, 14260-1500, USA}

\affiliation[b]{Department of Physics, University of Cincinnati, Cincinnati, Ohio 45221, USA}

\emailAdd{amartyas@buffalo.edu}
\emailAdd{ds77@buffalo.edu}
\emailAdd{rohana.wijewardhana@gmail.com}

\abstract{
The only two gauge symmetries which remain unbroken today are $SU(3)_c$ and $U(1)_{\rm EM}$. Both are crucial for our universe to appear the way it does and for our form of life to exist. Unless we are very special observers living at the very end of the cosmological symmetry-breaking chain, there is no reason to believe that these two symmetries will remain unbroken forever. In this paper, we investigate the cosmological observational signatures of a late-time $SU(3)_c$ symmetry breaking. We introduce a model with a new colored scalar field whose potential supports a first-order phase transition through nucleation of true-vacuum bubbles. 

We analyze three physically distinct particle-production mechanisms associated with the expanding bubble wall. The first is direct nonthermal particle production from vacuum mismatch across the accelerating wall, including both the scalar and massive-gluon sectors. The second is particle production from frictional dissipation into a shocked layer. For this channel we include the finite-temperature effective potential and show that, for our benchmark, the critical temperature is below the massive-gluon and scalar masses; therefore an equilibrated relativistic thermal bath of broken-phase massive particles would restore $SU(3)_c$. We consequently formulate a subcritical thermal freeze-out estimate below $T_c$, where massive broken-phase particles can still be produced but only with Boltzmann suppression. The third is a non-equilibrium wall--matter transition-radiation channel, in which ambient baryonic matter crossing the relativistic $SU(3)_c\to U(2)$ wall can radiate broken-phase massive gluons and scalar excitations. This channel does not require a thermal bath above $T_c$ and can continue after the wall reaches terminal velocity.

We then study the decays of the physical color-octet scalar and the massive gluons, and use \texttt{Pythia} to hadronize their decay products and determine the resulting photon and neutrino spectra. If the wall reaches a subluminal terminal velocity, these photons and neutrinos can arrive before the wall itself. The resulting high-energy photon and neutrino spectra constitute a long-range observational signature which, if ever observed, could be interpreted as a signal of cosmic doomsday.
}

\begin{document}
\maketitle
\tableofcontents

\section{Introduction}
It is believed that our universe in its history went through a series of phase transitions in which an original gauge symmetry group was broken either completely or down to a smaller subgroup. These processes can be considered violent since the whole structure of the universe gets rearranged, and the resulting new phase in the universe does not resemble much the previous one.  The only two gauge symmetries that remain unbroken today are $SU(3)_c$ and $U(1)_{\rm EM}$. However, there is no reason to believe that they will remain unbroken forever. Once they break, the universe will rearrange again, which will have very dramatic consequences for all the life forms in it, including ours. For example, if $SU(3)_c$ symmetry breaks, the gluons will become massive, and quarks will no longer be confined inside protons and neutrons (except perhaps if one of the subgroups remains unbroken and thus the corresponding gluons remain massless). It is hard to imagine what the universe will look like in this new phase, but it is certain that it will not be able to support a life similar to ours. Even if such a possibility is very remote, the seriousness of its consequences warrants a detailed study of the signature of such a phase transition.  The purpose of this paper is to study the astrophysical signature of the late-time $SU(3)_c$ symmetry breaking (the signature of $U(1)_{\rm EM}$ symmetry breaking will be studied elsewhere \cite{Sengupta:2026xlp}).

We will first write down the model supporting $SU(3)_c$ gauge symmetry breaking, which is still consistent with the current observational data. To avoid most of the astrophysical and collider constraints, we consider a first-order phase transition. Such a model inevitably contains a new massive colored scalar field responsible for the symmetry breaking. We then derive the dynamics of true-vacuum bubble propagation and discuss the production of particles in the background of the expanding wall.

There are several physically different ways in which particles can be produced by such a wall, and it is useful to keep them separate from the beginning. The first mechanism is direct production from vacuum mismatch. As the wall passes, the local vacuum changes from the unbroken $SU(3)_c$ phase to the broken $SU(3)_c\to U(2)$ phase. The corresponding nonadiabatic change of the field modes produces massive scalar quanta and massive gluons. This is a clean and controlled nonthermal source, but it is also exponentially sensitive to the wall acceleration and becomes inefficient once the wall approaches terminal motion.

The second mechanism is associated with frictional dissipation. Interactions between the wall and the surrounding medium drain kinetic energy from the wall and deposit it into a shocked layer. A naive treatment would convert this energy into a relativistic thermal population of massive broken-phase particles. However, this is not automatically consistent. The same finite-temperature corrections that heat the plasma can also restore the symmetry. We therefore compute the finite-temperature critical temperature for our benchmark and show that $T_c$ lies below both the massive-gluon and color-octet scalar masses. Thus a thermal bath with $T\gtrsim M_g,M_s$ would not be a broken-phase gas of massive gluons and scalar octets. A consistent broken-phase thermal estimate must be subcritical, $T<T_c$, where the heavy particles are nonrelativistic and Boltzmann suppressed. We formulate this as a subcritical freeze-out estimate with an efficiency factor that parametrizes the fraction of dissipated energy which passes through such a broken-phase thermal stage.

The third mechanism is non-equilibrium wall--matter transition radiation. Ambient matter crossing the relativistic color-breaking wall is highly boosted in the wall frame. Since the wall changes the gauge spectrum from unbroken $SU(3)_c$ to $U(2)$, the color-field configuration of an incoming quark or hadron is rearranged across the interface. The wall can absorb the longitudinal momentum mismatch and allow radiation of massive broken-phase gluons or color-octet scalar excitations. Unlike an equilibrated thermal bath, this process is controlled by the wall-frame beam energy rather than by a plasma temperature, and it can continue even after the wall reaches terminal velocity.

Eventually, we are interested in the long-range observable signature detectable on Earth. We therefore calculate the decay widths and branching ratios of the physical color-octet scalar and the massive gluons. We then use \texttt{Pythia 8} to hadronize their decay products and determine the resulting photon and neutrino spectra. The different production mechanisms mainly change the normalization of the parent-particle population, while the decay and hadronization phenomenology of the TeV-scale parent states is treated in a common way. If the wall is slowed to a subluminal terminal velocity, the resulting photons and neutrinos can reach us before the wall itself, providing a possible advance signal of the transition.

\section{Model for \texorpdfstring{$SU(3)_c$}{SU(3)c} gauge symmetry breaking }
\label{Model}
When considering a model for $SU(3)_c$ gauge symmetry breaking, we have two options: either a first- or second-order phase transition. A second-order phase transition is a smooth process in which the field responsible for the symmetry breaking is smoothly rolling down its own potential. If we are interested in the late-time phase transition, such a field must be very light, i.e. with a mass of the order of the current temperature in the universe ($\sim 10^{-4}$eV). Such a field will be very difficult to hide from the current collider constraints.  We therefore choose to study a first-order phase transition involving a colored scalar field with a mass above the current collider limits. In this scenario, we live in the false vacuum where $SU(3)_c$ gauge symmetry is not broken. Eventually we will tunnel into the true vacuum, however, if the lifetime of the false vacuum is longer than the age of the universe, no observational constraints (either cosmological, or collider) will be violated.  
Although observational constraints indicate that most of our universe is currently in the unbroken phase, they do not exclude the possibility that bubbles of the true vacuum already exist within our universe.

\smallskip
\noindent
Consider a Lagrangian density that is invariant under $SU(3)_c$ transformations:
\begin{equation}
{\cal L} = -\frac{1}{2}\text{Tr} F_{\mu \nu}F^{\mu \nu} + \frac{1}{2}\text{Tr}(D_\mu \Phi)(D^\mu \Phi) - V(\Phi),
\end{equation}
where the field strength and covariant derivative are
\begin{equation}
F_{\mu\nu}
\;=\;
\partial_\mu A_\nu - \partial_\nu A_\mu
\;+\;
i\,g\,[A_\mu,A_\nu],
\quad
D_\mu \Phi
\;=\;
\partial_\mu \Phi
\;+\;
i\,g\,[A_\mu,\Phi],
\quad
A_\mu = A_\mu^a T^a,
\end{equation}
where $\mu,\nu=0,1,2,3$, and repeated indices are summed with the Minkowski metric $\eta^{\mu\nu}.$
The scalar field $\Phi$ is taken to be in the adjoint representation of $SU(3)_c$. $D^\mu \Phi$ represents the covariant derivative of $\Phi$, and $g$ is the gauge coupling constant. We use fundamental-representation generators normalized as
\begin{equation}
\Tr(T^aT^b)=\frac12\delta^{ab}.
\end{equation}
With this convention the adjoint scalar matrix may be written as
\begin{equation}
\Phi=\sqrt{2}\,\phi^aT^a,
\end{equation}
where the eight real fields \(\phi^a\) are canonically normalized. Indeed,
\begin{equation}
\frac12\Tr(D_\mu\Phi D^\mu\Phi)
=
\frac12(D_\mu\phi^a)(D^\mu\phi^a).
\end{equation}
Equivalently, the factor of \(\sqrt2\) may be absorbed into the definition of the matrix field \(\Phi\). In what follows, however, the one-dimensional tunneling field \(\psi\) should always be understood as the canonically normalized field along the \(SU(3)_c\to U(2)\) direction, not as a raw diagonal matrix entry.

Varying \(\Phi\) leads to the equation of motion
\begin{equation}
D_\mu D^\mu \Phi + \frac{\partial V}{\partial \Phi}=0.
\end{equation}
The expanded form of the covariant derivative acting on \(\Phi\) is therefore
\begin{equation}
{
D_\mu D^\mu \Phi = \eta^{\mu \nu} D_\mu D_\nu \Phi = \eta^{\mu \nu} \left[ \partial_\mu\partial_\nu \Phi + i\,g\,\Bigl([\partial_\mu A_\nu,\Phi] + 2\,[A_\mu,\partial_\nu\Phi]\Bigr) - g^2\,[A_\mu,[A_\nu,\Phi]] \right].
}
\end{equation}

The trace-level potential may be written as
\begin{equation}
V(\Phi)
=
\frac{\mu_\Phi^2}{4}(\text{Tr}\Phi^2)
+
\frac{\lambda_{1\Phi}}{16}(\text{Tr}\Phi^2)^2
+
\frac{\lambda_{2\Phi}}{6}(\text{Tr}\Phi^3)
+
V_0.
\end{equation}
We assume the breaking pattern $SU(3)_c \to U(2)$, which leaves four gauge bosons massless and gives mass to the remaining four. In the scalar sector, the four Goldstone modes associated with the broken generators are eaten by the massive gluons, leaving four physical scalar degrees of freedom. A detailed classification of the vacuum alignments and the corresponding unbroken subgroups that can arise from this potential is presented in Appendix~\ref{App-A}, where we summarize the possible symmetry-breaking patterns of $SU(3)_c$. We require the potential to be bounded from below and set $\lambda_{1\Phi} > 0$. The term $\text{Tr}\Phi^4$ is not included separately since for a traceless $3\times3$ matrix it is proportional to $(\text{Tr}\Phi^2)^2$, as shown in Appendix~\ref{App-B}. A similar model was previously studied (see \cite{Stojkovic:2007dw,PhysRevD.19.1906}). The constant \(V_0\) is an overall shift of the potential and does not affect the tunneling profile, particle masses, or production estimates considered below.

\smallskip
\noindent
In a convenient diagonal representation, the \(SU(3)_c\to U(2)\) direction is proportional to
\begin{equation}
\Phi(q)
=
\begin{pmatrix}
q & 0 & 0\\
0 & q & 0\\
0 & 0 & -2q
\end{pmatrix}.
\end{equation}
The quantity \(q\) is the raw diagonal matrix entry. It is not canonically normalized. Along this direction,
\begin{equation}
\frac12\Tr(\partial_\mu\Phi\,\partial^\mu\Phi)
=
\frac12\Tr
\begin{pmatrix}
\partial_\mu q & 0 & 0\\
0 & \partial_\mu q & 0\\
0 & 0 & -2\partial_\mu q
\end{pmatrix}^2
=
3(\partial_\mu q)(\partial^\mu q).
\end{equation}
Thus the canonically normalized one-dimensional field is
\begin{equation}
\psi=\sqrt6\,q,
\qquad
q=\frac{\psi}{\sqrt6}.
\label{eq:canonical-psi-def}
\end{equation}
Equivalently, along the breaking direction one may write
\begin{equation}
\Phi(\psi)
=
\frac{\psi}{\sqrt6}
\begin{pmatrix}
1 & 0 & 0\\
0 & 1 & 0\\
0 & 0 & -2
\end{pmatrix}
=
\sqrt2\,\psi\,T_8,
\end{equation}
where \(T_8=\mathrm{diag}(1,1,-2)/(2\sqrt3)\). With this definition the kinetic term is
\begin{equation}
\frac12\Tr(\partial_\mu\Phi\,\partial^\mu\Phi)
=
\frac12(\partial_\mu\psi)(\partial^\mu\psi).
\end{equation}
Therefore, throughout the rest of the paper, \(\psi\) denotes the canonically normalized order-parameter field.

Substituting Eq.~\eqref{eq:canonical-psi-def} into the trace-level potential gives
\begin{equation}
V(\psi)
=
\frac14\mu_\Phi^2\psi^2
+
\frac{1}{16}\lambda_{1\Phi}\psi^4
-
\frac{\lambda_{2\Phi}}{6\sqrt6}\psi^3
+
V_0.
\end{equation}
It is convenient to absorb the projection and normalization factors into effective one-field parameters,
\begin{equation}
\mu^2\equiv\frac{\mu_\Phi^2}{6},
\qquad
\lambda_1\equiv\frac{\lambda_{1\Phi}}{36},
\qquad
\lambda_2\equiv\frac{\lambda_{2\Phi}}{6\sqrt6}.
\label{eq:effective-canonical-parameters}
\end{equation}
In terms of these effective parameters the canonically normalized potential becomes
\begin{equation}
V(\psi) = \frac{3}{2}\mu^2 \psi^2 + \frac{9}{4}\lambda_1 \psi^4 - \lambda_2 \psi^3 + V_0.
\label{eq:canonical-reduced-potential}
\end{equation}
This is the reduced potential used in the rest of the paper. The parameters \(\mu^2,\lambda_1,\lambda_2\) appearing below are therefore the effective parameters of the canonically normalized tunneling coordinate, not the unprojected trace-level coefficients.

Introducing new definitions
\begin{equation}
\psi_0 \equiv \frac{2}{9}\frac{\lambda_2}{\lambda_1}, \quad \epsilon_0 \equiv \lambda_1 - \frac{2\mu^2}{3\psi_0^2},
\end{equation}
allows a more convenient rewriting of the potential
\begin{equation}
V(\psi) = \frac{9}{4}\lambda_1 \psi^2(\psi - \psi_0)^2 - \frac{9}{4}\epsilon_0 \psi_0^2 \psi^2 + V_0.
\end{equation}

The parameter $\epsilon_0$ represents the degree of the fine-tuning in $V(\psi)$. If $\epsilon_0 = 0$, the potential has two degenerate minima at $\psi = 0$ and $\psi = \psi_0$, and the fine-tuning is absent. For a small and positive $\epsilon_0$, the global minimum shifts slightly to $\psi = \psi_0(1 + \epsilon_0/\lambda_1)$. Expanding about the shifted minimum $\psi=\psi_0(1+\epsilon_0/\lambda_1),$ one finds at leading order in $\epsilon_0$, the difference in energy densities between these two vacua is
\begin{equation}
\delta V \simeq \frac94 \epsilon_0 \psi_0^4.
\end{equation}
\begin{figure}[t]
\centering
 \includegraphics[width=0.65\linewidth]{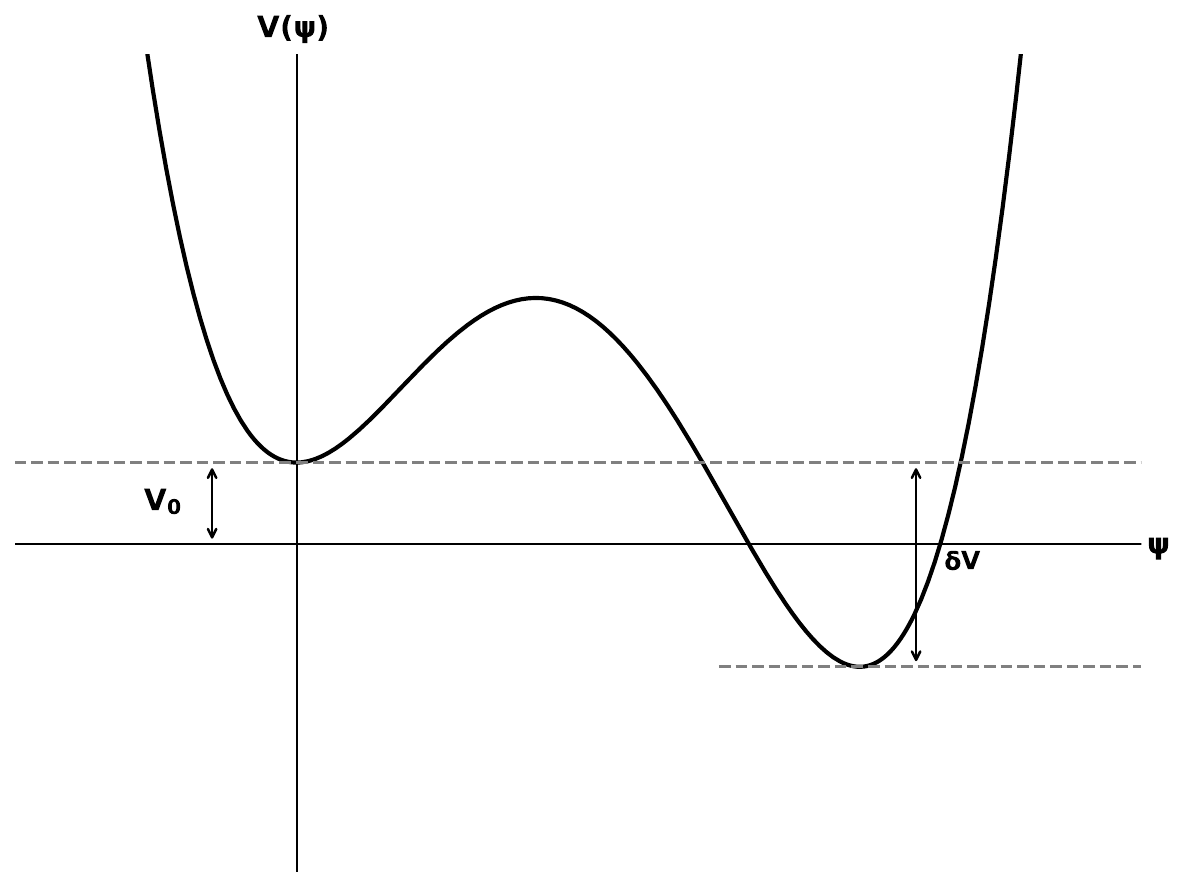}\\
 \caption{Characteristic potential for a first-order phase
 transition. There exist the false vacuum where the symmetry is
 unbroken, $\psi =0$, and the true vacuum where the symmetry is
 broken, $\psi \neq 0$. $\delta V$ is the difference in
 energy densities between the vacua. $V_0$ is the overall shift of the potential.} \label{Vphi} \end{figure}

Figure \ref{Vphi} illustrates the characteristic potential for this first-order phase transition scenario. It shows the false vacuum with unbroken symmetry at $\psi = 0$ and the true vacuum with broken symmetry at $\psi \neq 0$. The difference in energy density between these vacua is denoted by $\delta V$, while $V_0$ is the overall shift of the potential that our calculations do not depend on. But for example, we could use this freedom to shift the whole potential up to avoid tunneling into an AdS space with negative cosmological constant (where the whole region would collapse into a black hole).

The presence of a deeper true minimum in the potential indicates that any location in the universe will ultimately tunnel into the true vacuum, where the $SU(3)_c$ symmetry is broken. However, the rate of tunneling is determined by the energy difference between the vacua and the height of the potential barrier.
For a phenomenologically valid model, the tunneling rate must be sufficiently slow so that the lifetime of the false vacuum is at least of the order of the age of the universe. 

In the false vacuum state, where $\psi = 0$, excitations of the new colored scalar field have a mass scale $m_\psi \sim \mu$. A natural choice for $\mu$ is around ${\cal O}(1{\rm TeV})$, aligning with the anticipated scale for new physics beyond the Standard Model. Moreover, it would be challenging to hide a colored scalar field significantly lighter than this scale. Selecting this scale helps avoid the strongest existing experimental constraints, although a dedicated collider reinterpretation would be model-dependent.

Next, we examine the constraints on the parameter $\epsilon_0$. A crucial question is how long the universe could exist in the supercooled false vacuum state $\psi=0$. Transitioning from this false to the true  vacuum occurs through bubble nucleation of the true vacuum within the false vacuum. Within the semi-classical approximation, the transition probability per unit space-time volume is given by
\begin{equation} \label{prob}
\Gamma = {\cal A} e^{-S_E},
\end{equation}
where $S_E$ is the Euclidean action of the $O(4)$-symmetric bounce solution describing tunneling, and ${\cal A}$ is a constant of order ${\cal O}({\rm TeV}^4)$. Since we only consider the magnitude of the transition rate, we neglect the uncertainty in the prefactor. To compute $S_E$, we use the standard method defined in \cite{Coleman:1977py}.

At zeroth order in \(\epsilon_0\), the wall tension, or one-dimensional Euclidean action across the wall, is
\begin{equation}
S_1 = \int_{0}^{\psi_0} d\psi' \sqrt{2V(\psi')} \approx \sqrt{\frac{\lambda_1}{8}} \psi_0^3.
\end{equation}

In the thin-wall approximation and zero-temperature limit, the radius of the critical bubble is given by
\begin{equation}
R_0 = \frac{3S_1}{\delta V} = \frac{\sqrt{2\lambda_1}}{3} \frac{1}{\epsilon_0 \psi_0}.
\end{equation}

The Euclidean action for an $O(4)$-symmetric bubble is:
\begin{equation}
S_E = -\frac{1}{2}\delta V\pi^2 R_0^4 + 2\pi^2 R_0^3 S_1 = \frac{\pi^2 \lambda_1^2}{54}\frac{1}{\epsilon_0^3}.
\end{equation}

The decay rate per unit volume and time is defined in (\ref{prob}). We require that our observable universe, with a four-volume of the order of $t_{\rm Hubble}^4$, remains in the unbroken phase (i.e. false vacuum). This requirement corresponds to $\Gamma t_{\rm Hubble}^4 \lesssim 1$. With $t_{\rm Hubble} \sim 10^{10}$ years, this implies the limit $S_E > 400$. For a generic value $\lambda_1 \sim 1$, vacuum stability imposes only a mild fine-tuning condition, $\epsilon \lesssim 0.1$. Furthermore, $\epsilon_0 \lesssim 0.1$ is also sufficient to validate the thin-wall approximation. For this study, we adopt the benchmark values $\epsilon_{0}=0.076$ and $\lambda_{1}=1$, with a true vacuum expectation value of $\psi=1058~\mathrm{GeV}$. These choices yield $\psi_{0}=983.3~\mathrm{GeV}$, surface tension $S_{1}=3.36\times10^{8}~\mathrm{GeV}^{3}$, vacuum energy difference $\Delta V=1.60\times10^{11}~\mathrm{GeV}^{4}$, and a critical bubble radius $R_{0}=6.31\times10^{-3}~\mathrm{GeV}^{-1}$. The corresponding bounce action is $S_{E}\simeq 416$. \\
Evaluating the decay probability with ${\cal A}=(1~\mathrm{TeV})^4$, we obtain
\begin{equation}
\Gamma\,t_{\rm Hubble}^{4} \;=\; 7.96\times 10^{-3}\,,
\end{equation}
well below unity. With our chosen parameters corresponding to an $SU(3)_c$ breaking scale approximately of order $1~\mathrm{TeV}$, this represents the closest value attainable to unity within the phenomenologically allowed region of parameter space. This corresponds to a tunneling rate
\begin{equation}
\Gamma
\simeq
(1~{\rm TeV})^4 e^{-416.36}
\simeq
1.5\times10^{-181}~{\rm TeV}^4
\simeq
1.5\times10^{-169}~{\rm GeV}^4,
\end{equation}
which ensures that the false vacuum at $\psi=0$ is cosmologically long-lived. Thus, the $SU(3)_c$ false vacuum is metastable on Hubble timescales, satisfying the stability criterion $\Gamma t_{\rm Hubble}^4 \lesssim 1$ (see Section~\ref{ParticleProduction}).

\section{Propagation of the true vacuum bubble}

When the scalar field tunnels through the potential barrier, a bubble of true vacuum forms and begins to expand \cite{Coleman:1977py,Dai:2019eei,Sengupta:2025jah}. To get the dynamics of the bubble, we neglect the interactions between $\Phi$ and other fields (e.g. fermions, gauge fields, etc.). In this section \(\Phi\) denotes the canonically normalized tunneling profile along the bounce direction. The dynamics of the scalar field, $\Phi (\rho)$, is then given by the action 
\begin{equation}
S(\Phi) = \int d^4x \left(\frac{1}{2}(\partial_\mu \Phi)^2 - V(\Phi)\right),
\end{equation}
yielding the classical equation of motion
\begin{equation}
-\partial_t^2 \Phi + \nabla^2 \Phi - V'(\Phi) = 0.
\end{equation}
Upon Wick rotation, $t \rightarrow i\tilde{\tau}$, we obtain the following
\begin{equation}
\partial_{\tilde{\tau}}^2 \Phi + \nabla^2 \Phi - V'(\Phi) = 0.
\end{equation}
Considering a $O(4)$ symmetric bounce solution, the equation further simplifies to
\begin{equation}
\frac{d^2\Phi}{d\rho^2} + \frac{3}{\rho}\frac{d\Phi}{d\rho} = V'(\Phi),
\end{equation}
with $\rho = \sqrt{\tilde{\tau}^2 + r^2}$, indicating dependence only on $\rho$. The bounce solution satisfying the boundary conditions at nucleation ($t=0$) is
\begin{equation}
\Phi(t,\vec{x}) = \Phi\bigl(\rho=\sqrt{r^2-t^2}\bigr).
\end{equation}
Here, the bubble contracts for $t<0$, bounces at $t=0$, and expands afterward. As usual, we cut the solution at $t=0$ and consider only the expansion. Utilizing the thin-wall approximation, we express the scalar-field configuration as
\begin{equation}
\Phi(\rho) = \begin{cases}
v, & \rho > R,\\[1ex]
v_1, & \rho < R,
\end{cases}
\end{equation}
where $v$ and $v_1$ represent the scalar field expectation values in false and true vacua, respectively, while $R$ is the radius of the bubble.

Particle production during first‐order phase transitions has been extensively studied in the literature (e.g., \cite{Sengupta:2025jah,Yamamoto:1994te,MersiniHoughton:1999tt,Mersini-Houghton:1999aoa,Maziashvili:2003kp,Espinosa:2025ejf,Isidori:2001bm,Hiller:2024zjp,Espinosa:2023oml,Espinosa:2016nld,Alonso:2023jsi,Maziashvili:2003sk,Maziashvili:2003zy,Vachaspati:1991tq,Swanson:1986hx,Hamazaki:1995dy,Maziashvili:2003kj,Kobzarev:1974cp, Hogan:1983ixn, Kamionkowski:1993fg,Baker:2021nyl,Baker:2021sno,Witten:1984rs, Shakya:2023kjf,Kolb:2023dzp,Kolb:2023ydq,Garcia-Bellido:2001dqy,Espinosa:2015qea,Linde:1981zj,Dine:1992wr,Linde:1980tt,Linde:1977mm,Dine:1992vs,Kallosh:2003mt,Kallosh:2003bq,Krive:1976sg,Bentivegna:2017qry,Kawana:2022lba,Degrassi:2012ry,Turner:1992tz,Quiros:1999jp,Steinhardt:1981ct,Espinosa:2010hh,Ai:2023yce}).
For our purposes, we begin by employing the vacuum-mismatch method, following Refs.~\cite{Maziashvili:2003zy,Tanaka:1993ez}, and later discuss the thermal production mechanism generated by frictional dissipation in the shocked layer behind the bubble wall.

\section{Particle-production mechanisms considered in this work}
\label{sec:production-overview}

Before turning to the detailed calculation, we summarize the particle-production mechanisms considered in this work. The expanding $SU(3)_c$-breaking wall can produce particles in three conceptually distinct ways.

The first mechanism is vacuum-mismatch production. In this channel, particles are produced because the quantum fields experience different masses and mode functions on the two sides of the wall. The wall changes the local vacuum from the unbroken $SU(3)_c$ phase to the broken $SU(3)_c\to U(2)$ phase. This produces massive scalar excitations and massive gluons through nonadiabatic mode matching. This mechanism is nonthermal in the sense that it does not depend on the environmental temperature, but it is controlled by the magnitude of the wall acceleration which in turn determines the Unruh temperature of the radiation. It gives a conservative baseline signal, but it becomes strongly suppressed once the wall approaches terminal motion and the wall acceleration ceases

The second mechanism is thermal production from frictional dissipation. The same interactions which slow the wall also deposit energy into a shocked layer near the wall. If this energy thermalizes while the local temperature remains below the finite-temperature restoration scale, $T<T_c$, then massive broken-phase gluons and scalar octets can be produced from the thermal tail. Here \(M_g\) denotes the mass of the broken-phase massive gluons, and \(M_s\) denotes the physical color-octet scalar mass. Since our benchmark satisfies
\begin{equation}
T_c<M_g<M_s,
\end{equation}
these particles are necessarily nonrelativistic in the broken phase and their abundance is Boltzmann suppressed. We therefore treat this contribution as a subcritical freeze-out estimate, parametrized by an efficiency factor $f_{\rm th}$. If the shocked layer instead equilibrates above $T_c$, the local phase is restored and the broken-phase massive-particle description no longer applies.

The third mechanism is non-equilibrium wall--matter transition radiation. Ambient baryons crossing the relativistic wall appear as a high-energy beam in the wall frame. Because the gauge spectrum changes across the $SU(3)_c\to U(2)$ interface, quarks and hadrons crossing the wall can radiate massive broken-phase gluons or color-octet scalars. This process does not require a thermal bath above $T_c$ and can continue even after the wall has reached terminal velocity. Its total yield is therefore controlled by the amount of matter swept up by the wall and by microscopic radiation efficiencies.

In all three cases, once the heavy parent particles are produced, their subsequent decays and hadronization are described by the same particle-physics phenomenology. The massive gluons and color-octet scalars decay into quarks and gluons, which then hadronize and cascade into photons, neutrinos, and other stable particles. The three production mechanisms therefore differ mainly in how many parent particles they produce. Since the wall--matter transition-radiation channel gives the largest parent-particle yield for the benchmark parameters considered here, in the phenomenological plots below we focus on the photon and neutrino spectra generated by this dominant production channel.

\section{Particle production due to vacuum mismatch}
\label{ParticleProduction}
The difference between the false and true vacua typically leads to particle creation. Outside, the bubble remains the false vacuum, while inside the field transitions to the true vacuum via tunneling. This transition initiates the generation of massive scalar particles. By separating the scalar field into background and fluctuations, $\Phi = \Phi_c + h$, the fluctuation field $h$ obeys
\begin{equation}
\partial_{\tilde{\tau}}^2 h + \nabla^2 h - V''(\Phi_c)h = 0.
\end{equation}
Neglecting additional field interactions, this equation approximates as
\begin{eqnarray}
\partial_{\tilde{\tau}}^2 h + \nabla^2 h - M^2 h &=& 0,\quad \tilde{\tau} < \tilde{\tau}^*,\\[1ex]
\partial_{\tilde{\tau}}^2 h + \nabla^2 h - \mu^2 h &=& 0,\quad \tilde{\tau} > \tilde{\tau}^*,
\end{eqnarray}
where $\tilde{\tau}^*$ is the characteristic (Euclidean) time scale for the duration of the phase transition. We take here $\tilde{\tau}^*=-R_0$, where $R_0$ is the size of the bubble at the time of nucleation. The reason for this is that the bubble propagation is a motion with a constant proper (in contrast to coordinate) acceleration. From the equation of motion $r^2-t^2=R_0^2$, one can derive the magnitude of the proper acceleration as $a=1/R_0$ (see appendix). Thus, the process of particle creation in this context admits an Unruh-like interpretation, in which the amount of radiation is determined by the proper acceleration. Since in the absence of friction this acceleration is constant during the bubble expansion, particles are produced continuously as the bubble expands. We also note that the proper acceleration is defined in terms of the proper time of an observer who is momentarily at rest with respect to the bubble wall, and it is not equal to a coordinate acceleration measured by a distant observer (which actually goes to zero as the bubble wall approaches the speed of light).

The solution for $h$ can be written as a combination of mode functions $g_k$, which satisfy $\nabla^2 g_k=-k^2$
\begin{eqnarray}
g_k=\left\{
  \begin{array}{lr}
   e^{\omega_- {\tilde{\tau}}} e^{i\vec{k}\cdot \vec{x}} &  \text{, for ${\tilde{\tau}}<\tilde{\tau}^*$}\\
     A_k e^{\omega_+ {\tilde{\tau}}} e^{i\vec{k}\cdot \vec{x}} +B_k e^{-\omega_+ {\tilde{\tau}}} e^{i\vec{k}\cdot \vec{x}} & \text{, for ${\tilde{\tau}}>\tilde{\tau}^*$}
  \end{array}
\right.
\end{eqnarray}
For a scalar mode of momentum $k$, we use
\begin{equation}
\omega_+(k)=\sqrt{k^2+M_s^2},
\qquad
\omega_-(k)=\sqrt{k^2+m_{sf}^2},
\end{equation}
where $+$ and $-$ denote the true- and false-vacuum regions respectively. For the massive-gluon transverse modes we use
\begin{equation}
\omega_+(k)=\sqrt{k^2+M_g^2},
\qquad
\omega_-(k)=|k|.
\end{equation}
At $k=0$, these reduce to the expressions used in the zero-mode estimates below. Concretely, the scalar mass in the false vacuum region is
\begin{equation}\label{Mass_False_Vacuum}
{
M^{2}
=
\frac{9}{2}\,\psi_{0}^{2}\,\bigl[\lambda_{1}-\epsilon_{0}\bigr],
}
\end{equation} while in the true vacuum region is
\begin{equation}\label{Mass_True_Vacuum}
\mu^{2}
\;=\;
\frac{9}{2}\,\psi_{0}^{2}
\left[
\lambda_{1}
\;+\;
5\,\epsilon_{0}
\;+\;
6\,\frac{\epsilon_{0}^{2}}{\lambda_{1}}
\right].
\end{equation}
Since $g_k$ and $\partial_{\tilde{\tau}} g_k$ must be continuous at $\tilde{\tau} =\tilde{\tau}^*$, for $A_k$ and $B_k$ we get
\begin{eqnarray}
A_k&=&\frac{1}{2\omega_+}(\omega_++\omega_-)e^{-(\omega_+-\omega_-)\tilde{\tau}^*}\\
B_k&=&\frac{1}{2\omega_+}(\omega_+-\omega_-)e^{(\omega_++\omega_-)\tilde{\tau}^*} .
\end{eqnarray}
As explained above, we set $\tilde{\tau}^*=-R_0$. The particle creation spectrum is obtained from the Bogoliubov transform \cite{Tanaka:1993ez}
\begin{eqnarray}
N_k=\frac{B_k^2}{A_k^2-B_k^2} =  \left[ \frac{(\omega_++\omega_-)^2}{(\omega_+-\omega_-)^2}e^{4\omega_+ R_0}- 1 \right]^{-1}.
\label{pn}
\end{eqnarray}
The quantity $N_k$ is the occupation number of the momentum mode $k$. A physical number density is obtained by integrating over phase space, $d^3k/(2\pi)^3$. In the numerical estimates below we retain only the zero-mode contribution and use a fixed effective zero-mode normalization as a conservative proxy for the production density. A full phase-space integration over higher momentum modes would increase the total yield.
\begin{figure}[t]
\begin{center}
\includegraphics[scale=0.42]{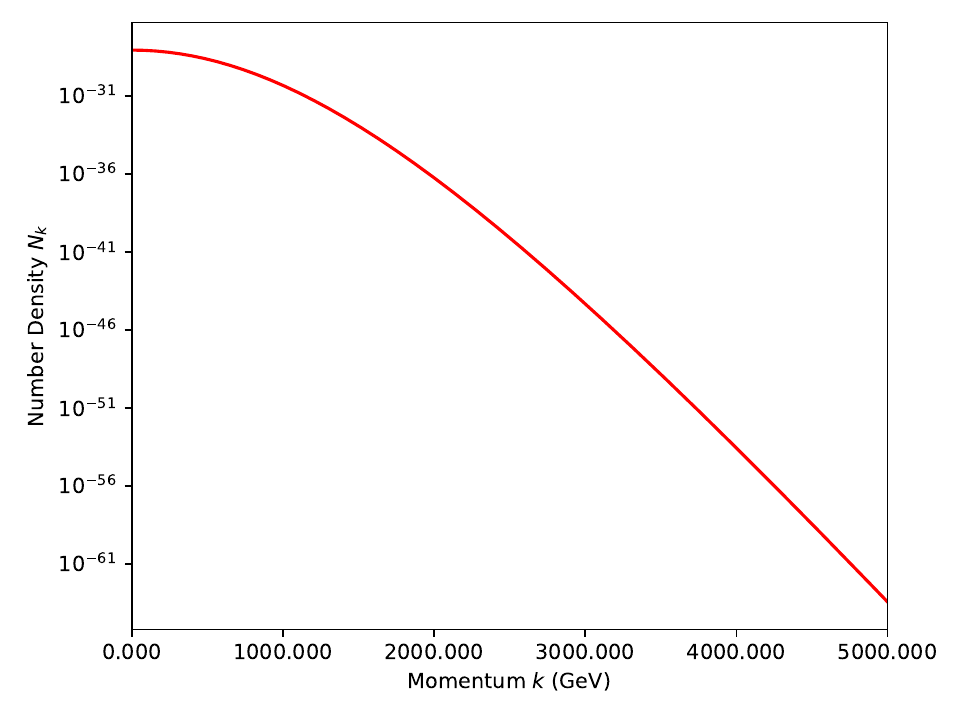}
\includegraphics[scale=0.42]{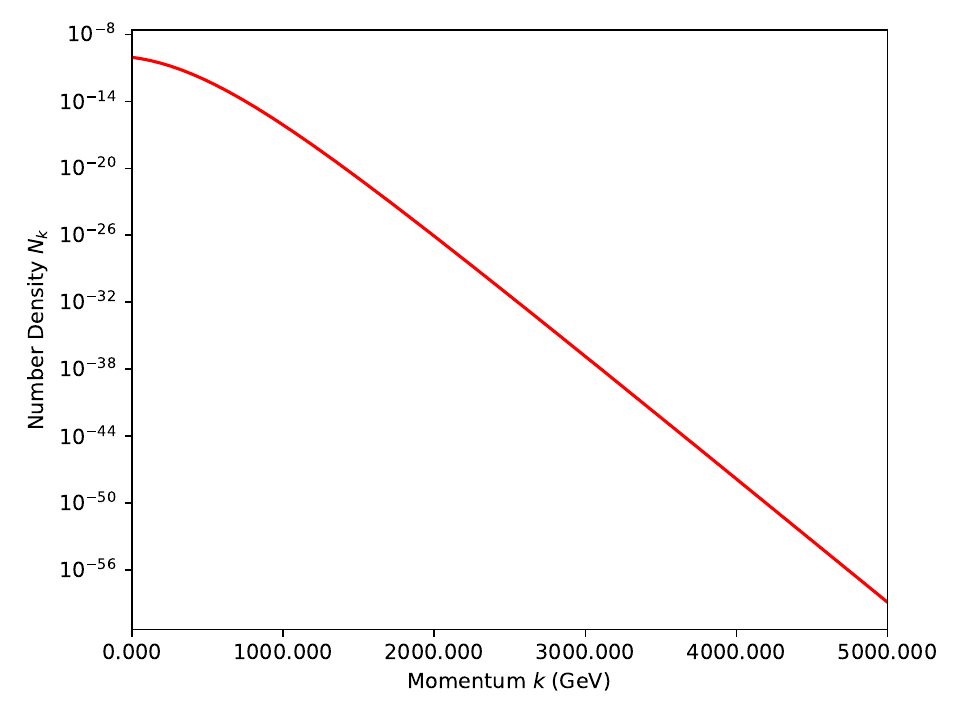}
\end{center}
\caption{\label{NDPlot}On the left we have the mode occupation $N_k$ of the scalar colored particles and on the right we have the mode occupation $N_k$ of the massive gluons created as a function of their momenta due to the vacuum mismatch.}
\end{figure}

 The left-hand side of Figure~\ref{NDPlot} presents the momentum-dependent mode occupation $N_k$ for the scalar field $\Phi$. For our chosen parameters discussed previously, i.e. $\epsilon_{0} = 0.076$, $\lambda_{1} = 1$, we get the scalar mass in the false vacuum to be $m_{sf} = 2000$ GeV,\,the bubble radius at the moment of nucleation is $R_0 \approx 6.3\, \text{TeV}^{-1}$, while the scalar mass in the true vacuum is $M_s \sim 2500$ GeV for the true vacuum located at $1058$ GeV. The large vacuum pressure released during bubble expansion $(\Delta V \sim 0.16~\mathrm{TeV}^4)$ provides the main energy source for producing massive particles in the true vacuum phase. The details of this production mechanism will be discussed later.

In models of this type, where the QCD gauge symmetry $SU(3)_c$ is broken by an adjoint scalar at the TeV scale, the residual unbroken subgroup is generically $U(2)$ (see Appendix~\ref{App-A}). In this vacuum alignment four gluons remain massless while the other four acquire masses
\begin{equation}
M_g \simeq \frac{\sqrt3}{2}\,g_s\,\psi_{\rm true},
\end{equation}
which is numerically of order \(1~\mathrm{TeV}\) for the benchmark used here, where \(\psi_{\rm true}\simeq1058~\mathrm{GeV}\) is the shifted true-vacuum expectation value. Adopting unitary (Proca) gauge, the massive gluons absorb the corresponding colored Goldstone modes and propagate as massive vectors.

The analysis of gluon production at zero comoving momentum requires additional care. Inside the broken phase, four of the eight gluons acquire mass, behaving as massive Proca fields with two transverse and one longitudinal polarization each, while the remaining four gluons stay massless. Before symmetry breaking, however, only the two transverse gluonic degrees of freedom are present as physical propagating modes; the longitudinal polarization emerges only after the vacuum expectation value of the adjoint scalar is established, when the corresponding colored Goldstones are absorbed. This distinction has important consequences for the mode matching at the bubble wall.

For the \emph{transverse} polarizations, the matching proceeds in a close analogy with the photon calculation. At $k=0$, one has $\omega_- = |k| = 0$ in the false vacuum and $\omega_+ = \sqrt{k^2+M_g^2}=M_g$ in the true vacuum. The Bogoliubov transformation then gives
\begin{equation}
N^{(T)}_{k=0} \;=\; \Bigl[e^{\,4 M_g R_0}-1\Bigr]^{-1},
\end{equation}
which is well defined for each transverse polarization. Including the multiplicity of the four massive gluons, each with two transverse polarizations, the transverse-only contribution to the gluon number density becomes
\begin{equation}
n_{\text{gluon}}(k=0)\;=\;8\,N^{(T)}_{k=0}
\;=\;\frac{8}{e^{\,4 M_g R_0}-1}\,.
\end{equation}

The \emph{longitudinal} gluon polarization, in contrast, cannot be captured by this simple quench calculation. Since it is absent before the symmetry is broken, one cannot directly apply the two-oscillator matching used for the transverse modes. A correct treatment requires setting up the coupled system of the gauge bosons and the colored Goldstone fields, $(G_\mu^a, \pi^a)$, and working in a covariant gauge such as $R_\xi$. The relevant gauge-invariant combinations, e.g.
\begin{equation}
\partial_\mu \pi^a - g_s v_c\, G_\mu^a \,,
\end{equation}
must then be matched consistently across the wall. The resulting Bogoliubov coefficients for the longitudinal modes are in general different from those of the transverse ones, and will depend sensitively on the details of the scalar sector and the choice of gauge fixing. A full computation of this coupled scalar–vector dynamics is beyond the scope of the present analysis and will be developed in future work. For the purposes of the present phenomenological study, we therefore adopt the conservative choice of retaining only the transverse polarizations when quoting gluon production rates and spectra.

We emphasize that this estimate undercounts the full gluonic production, since the longitudinal polarizations of the massive gluons are omitted. However, this omission does not qualitatively affect our cosmological conclusions, since including the longitudinal modes would change the total massive-gluon multiplicity only by an $\mathcal{O}(1)$ factor rather than by an order of magnitude, and is therefore negligible on the logarithmic scales relevant for the present signal estimates. Nevertheless, adopting the transverse-only contribution provides a conservative lower bound on the gluon yield relevant for the present phenomenological study. The corresponding spectra are shown in the right-hand panel of Fig.~\ref{NDPlot}, and the derivation of the transverse mode functions used in this expression is summarized in Appendix~\ref{App-C}.

\section{Relativistic bubble--wall dynamics in a viscous medium and terminal velocity}
\label{sec:bubble-dynamics-su3}

If the bubble wall moves with the speed of light, then no signal can overtake it to warn us
about the impending doom. However, expansion of a vacuum bubble is generically impeded by interactions with its environment, i.e. matter or radiation (where some of it could be produced by the bubble itself). The wall therefore tends to approach a subluminal terminal speed rather than accelerating indefinitely. In this work we use the relativistic thin--wall framework developed in our earlier analysis of Higgs--vacuum decay with friction~\cite{Si:2025vdt,Sengupta:2025jah}, and we adapt it to the present SU(3)\(_c\)–breaking setting (see also~\cite{Moore_2000, Moore:1995ua, Branchina:2025jou, Bodeker:2017cim, Azatov:2020ufh, Azatov:2023xem,Ai:2025bjw,Kobzarev:1974cp,Giombi:2023jqq,Cai:2020djd,Balaji:2020yrx,Ai:2021kak,Ai:2023see} for related discussions of wall friction and hydrodynamics). For completeness, we collect the working equations needed below and record the SU(3)\(_c\) model–specific normalizations.

\subsection{Equation of motion and terminal balance}
\label{subsec:eom-su3}

In the thin--wall description the rest energy per unit area equals the surface tension \(\sigma\). Under a boost with speed \(v\) the wall carries energy density \(\sigma\gamma(v)\) and momentum density \(\sigma\gamma(v)\,v\), with \(\gamma(v)=(1-v^2)^{-1/2}\). Balancing the time derivative of the momentum density against the net outward force density (vacuum pressure jump, curvature pressure, and linear drag) gives the relativistic spherical EOM~\cite{Sengupta:2025jah}:
\begin{equation}
  \sigma\,\gamma^3(v)\,\frac{dv}{dt}
  \;=\;
  \Delta V \;-\; \frac{2\sigma}{R(t)} \;-\; \eta\,\gamma(v)\,v,
  \label{eq:spherical_EOM_su3}
\end{equation}
where \(R(t)\) is the instantaneous radius, \(\Delta V\equiv V_{\rm false}-V_{\rm true}\) is the driving pressure (latent heat), and \(\eta\) is the effective linear friction coefficient encoding wall–medium interactions.

At terminal motion the acceleration vanishes and Eq.~\eqref{eq:spherical_EOM_su3} reduces to
\begin{equation}
  0 \;=\; \Delta V \;-\; \frac{2\sigma}{R} \;-\; \eta\,\gamma(v)\,v
  \qquad\Longleftrightarrow\qquad
  \Delta V_{\rm eff}(R)\;\equiv\;\Delta V-\frac{2\sigma}{R}
  \;=\; \eta\,\gamma(v)\,v,
  \label{eq:terminal_balance_su3}
\end{equation}
so in the large–\(R\) (planar) limit the terminal relation is simply \(\Delta V=\eta\,\gamma v\).

\medskip
For the SU(3)\(_c\) potential specified in Sec.~\ref{Model}, the quantities that enter Eq.~\eqref{eq:spherical_EOM_su3} are
\begin{equation}
  \Delta V \;=\; \frac{9}{4}\,\epsilon_0\,\psi_0^4,
  \qquad
  \sigma \;\simeq\; \sqrt{\frac{\lambda_1}{8}}\;\psi_0^3,
\end{equation}
so that
\begin{equation}
  A \;\equiv\; \frac{\Delta V}{\sigma}
  \;=\; \frac{3}{R_0},
  \label{eq:A_equals_3_over_R0_su3}
\end{equation}
with \(R_0\) the thin--wall nucleation radius. Equation~\eqref{eq:A_equals_3_over_R0_su3} fixes the initial drive used below.

\subsection{Proper--time formulation}
\label{subsec:proper-su3}

Following~\cite{Sengupta:2025jah}, it is convenient to parametrize the motion by the wall’s proper time \(\tau\) and rapidity \(y(\tau)\),
\begin{equation}
  v=\tanh y,\qquad
  \gamma=\cosh y,\qquad
  \gamma v=\sinh y,\qquad
  \frac{dt}{d\tau}=\gamma,\qquad
  \frac{dR}{d\tau}=\gamma v=\sinh y.
  \label{eq:kinematics_su3}
\end{equation}
The proper acceleration is
\begin{equation}
  \alpha(\tau)\;\equiv\;\gamma^3\,\frac{dv}{dt}\;=\;\frac{dy}{d\tau}.
  \label{eq:alpha_def_su3}
\end{equation}
Dividing Eq.~\eqref{eq:spherical_EOM_su3} by \(\sigma\) and using \eqref{eq:alpha_def_su3} yields
\begin{align}
  \frac{dy}{d\tau}
  &= A \;-\; \frac{2}{R(\tau)} \;-\; B\,\sinh y(\tau),
  \qquad
  A\equiv\frac{\Delta V}{\sigma},\ \ B\equiv\frac{\eta}{\sigma},
  \label{eq:dy_dtau_su3}\\
  \frac{dt}{d\tau}&=\cosh y,\qquad
  \frac{dR}{d\tau}=\sinh y.
  \label{eq:dt_dR_dtau_su3}
\end{align}
With nucleation data \(R(0)=R_0\) and \(y(0)=0\), one has
\begin{equation}
  \alpha(0)=A-\frac{2}{R_0}\;\simeq\;\frac{1}{R_0},
  \label{eq:alpha0_su3}
\end{equation}
where the final step uses \(A=3/R_0\) from Eq.~\eqref{eq:A_equals_3_over_R0_su3}.\\
For proper times \(\tau\ll R_0\) the motion is only mildly relativistic. With
\(y(\tau)=\alpha(0)\tau+\mathcal{O}(\tau^2)\) and \(\sinh y\simeq y\) one finds
\begin{equation}
  \sinh y(\tau)\simeq \alpha(0)\,\tau,
  \qquad
  R(\tau)\simeq R_0+\int_0^\tau \sinh y\,d\tau'
  \simeq R_0+\frac{\alpha(0)}{2}\tau^2.
  \label{eq:early_kin_su3}
\end{equation}
Inserting \eqref{eq:early_kin_su3} into \(\alpha(\tau)=A-\tfrac{2}{R(\tau)}-B\,\sinh y(\tau)\)
[Eq\ \eqref{eq:dy_dtau_su3}] gives the leading drag and curvature pieces
\(-B\,\alpha(0)\tau\) and \(+\alpha(0)\tau^2/R_0^2\), so the proper acceleration expands as
\begin{equation}
  \alpha(\tau)\simeq
  \underbrace{\Big(A-\frac{2}{R_0}\Big)}_{\alpha(0)}
  \;-\;B\,\alpha(0)\,\tau
  \;+\;\frac{\alpha(0)}{R_0^2}\,\tau^2
  \;+\;\mathcal{O}(\tau^3).
  \label{eq:alpha_series_su3}
\end{equation}
Using \(A\simeq 3/R_0\) implies \(\alpha(0)\simeq 1/R_0\), yielding
\begin{equation}
  \alpha(\tau)\simeq
  \frac{1}{R_0}-\frac{B}{R_0}\tau+\frac{\tau^2}{R_0^3}
  +\mathcal{O}(\tau^3),
  \label{eq:alpha_series_simplified_su3}
\end{equation}
valid for \(\tau\ll R_0\) (so \(y\ll 1\)) and weak friction in the sense \(B\,R_0\ll 1\).
The linear term \(-\,(B/R_0)\tau\) arises from viscous drag, whereas the quadratic term \(+\tau^2/R_0^3\) reflects the relaxation of the contribution to curvature as \(R\) grows.
\subsection{Evolution of the proper acceleration and approach to terminal motion}
\label{subsec:alpha-evolution-su3}

Differentiating Eq.~\eqref{eq:dy_dtau_su3} and using \(\frac{dR}{d\tau}=\sinh y\) and \(\frac{dy}{d\tau}=\alpha\) yields the exact evolution equation
\begin{equation}
  \frac{d\alpha}{d\tau}
  \;=\;\frac{2}{R^2}\,\frac{dR}{d\tau}\;-\;B\,\cosh y\,\frac{dy}{d\tau}
  \;=\;\frac{2\,\sinh y}{R^2}\;-\;B\,\cosh y\,\alpha.
  \label{eq:alpha_prime_su3}
\end{equation}
In the planar, small–rapidity limit (\(R\to\infty\), \(\cosh y\simeq 1\)) one has \(d\alpha/d\tau\simeq -B\,\alpha\), hence
\begin{equation}
  \alpha(\tau)\;\simeq\;\alpha(0)\,e^{-B\,\tau},
  \qquad
  \tau_{\rm term}\;\equiv\;\frac{1}{B}\;=\;\frac{\sigma}{\eta},
  \label{eq:tau_term_su3}
\end{equation}
which defines the characteristic proper--time to approach terminal motion. As $R\to\infty$ the curvature term disappears and \eqref{eq:dt_dR_dtau_su3} reduces to
\begin{equation}
\alpha(\tau)\;\to\;A-B\,\sinh y(\tau).
\label{eq:alpha-planar-limit}
\end{equation}
At terminal balance (\(\alpha\to 0\)) the rapidity satisfies
\begin{equation}
  \sinh y_{\rm term}=\frac{A}{B}
  \qquad\Longleftrightarrow\qquad
  \gamma_{\rm term}\,v_{\rm term}=\frac{\Delta V}{\eta},
  \label{eq:terminal_sinh_su3}
\end{equation}
the planar version of Eq.~\eqref{eq:terminal_balance_su3}. These relations, taken together with the SU(3)\(_c\) identification \(A=3/R_0\), are the inputs we use to benchmark wall velocities and to delimit the acceleration–driven particle–production epoch in the remainder of this work.

\section{Particle production with frictional effects due to vacuum mismatch}
\label{sec:su3c-prod-spherical}

We now turn to the $SU(3)_c$ model, where the expansion of the bubble wall in the broken phase produces massive gluons and associated colored scalars. Our goal is to estimate the number of quanta produced in the regime prior to saturation at terminal velocity. For clarity we work here in TeV units. The relevant parameters determined by the potential in Sec.~\ref{Model} are
\begin{equation}\nonumber
\psi_0 \;=\; 0.983271 \;\text{TeV},
\qquad
\Delta V \;=\; 1.598416\times10^{-1}\;\text{TeV}^4,
\qquad
\sigma \;=\; 3.361052\times10^{-1}\;\text{TeV}^3,
\end{equation}
\begin{equation}
R_0 \;=\; 6.308219\;\text{TeV}^{-1},
\qquad
A \;=\; 0.475570\;\text{TeV},
\qquad
M_s \;=\; 2.480876\;\text{TeV}.
\label{eq:inputs-su3c}
\end{equation}
The dimensionless drive and friction ratios are defined analogously,
\begin{equation}
A\equiv\frac{\Delta V}{\sigma},\qquad
B\equiv\frac{\eta}{\sigma},
\end{equation}
and the bubble wall kinematics are parametrized by its proper time $\tau$ and rapidity $y(\tau)$,
\begin{equation}
v=\tanh y,\quad
\gamma=\cosh y,\quad
\gamma v=\sinh y,\qquad
\frac{dt}{d\tau}=\gamma,\quad
\frac{dR}{d\tau}=\sinh y.
\label{Parameters-su3c}
\end{equation}
The proper acceleration that governs particle production is
\begin{equation}
\alpha(\tau)\;\equiv\;\frac{dy}{d\tau}
\;=\;A\;-\;\frac{2}{R(\tau)}\;-\;B\,\sinh y(\tau),
\label{eq:alpha-spherical-su3c}
\end{equation}
with initial conditions
\begin{equation}
R(0)=R_0,\qquad y(0)=0,\qquad t(0)=0,\qquad N_{\rm tot}(0)=0.
\end{equation}
$\alpha(0)=A-2/R_0\simeq 1/R_0$ under the thin-wall relation $A\simeq 3/R_0$.

The instantaneous zero momentum mode occupancy at a proper time $\tau$ is taken from the vacuum-mismatch procedure derived in the previous sections by replacing the constant proper acceleration with the time dependent one from this section
\begin{equation}
N_{k=0}(\tau)
=\left[\frac{(\omega_++\omega_-)^2}{(\omega_+-\omega_-)^2}
\exp\!\Bigl(\tfrac{4\omega_+}{\alpha(\tau)}\Bigr)-1\right]^{-1},
\label{eq:Nk0-alpha-su3c}
\end{equation}
with $\omega_+=M_s$ and $\omega_-=m_{sf}$, while for the massive-gluon channel we use $\omega_+=M_g$ and $\omega_-=0$ in the true and false vacuum respectively. The total number evolves as
\begin{equation}
\frac{dN_{\rm tot}}{d\tau}
=\mathcal{N}_0\,
N_{k=0}(\tau)\,
4\pi R(\tau)^2\sinh y(\tau),
\label{eq:dNtot}
\end{equation}
and is accumulated numerically until the wall reaches its terminal velocity, $\tau=\tau_{\rm term}=\sigma/\eta$. $N_{k=0}^{\rm(int)}$ denotes the accumulated yield obtained by integrating the production rate from nucleation up to time $\tau.$ Thus at $\tau=0,$ no time or volume has yet been accumulated, so $N_{k=0}^{\rm(int)}(0)=0.$ By contrast, $N_{k=0}(0)$ is an instantaneous occupation; it contributes to the total only after integrating over a finite interval and swept volume with the effective zero-mode normalization described above. Here \(\mathcal{N}_0\) is an effective zero-mode phase-space normalization with dimensions of \({\rm TeV}^3\). In the numerical estimates below we set \(\mathcal{N}_0=1~{\rm TeV}^3\). Thus the quoted vacuum-mismatch yields should
be interpreted as zero-mode-normalized benchmark yields rather than full phase-space-integrated particle numbers. The details of the numerical calculation have been described in Appendix~\ref{App-I} and here we list the results in the following Table~\ref{tab:su3c-yields-gluons}

\begin{table*}[tbhp]
\centering
\small
\begin{tabular}{@{}lccccc@{}}
\toprule
Scenario 
& $\eta\,[{\rm TeV}^4]$ 
& $\tau_{\rm term}\,[{\rm TeV}^{-1}]$ 
& $R_{\rm fin}\,[{\rm TeV}^{-1}]$ 
& Scalars--$N_{\rm tot}^{\rm(int)}$ 
& Massive Gluons--$N_{\rm tot}^{\rm(int)}$ 
\\
\midrule
$\delta=10^{-12}$ 
& $2.260\times10^{-7}$ 
& $1.487\times10^{6}$ 
& $1.051\times10^{12}$ 
& $6.87\times10^{4}$ 
& $1.42\times10^{14}$ 
\\
$\delta=10^{-11}$ 
& $7.148\times10^{-7}$ 
& $4.702\times10^{5}$ 
& $1.051\times10^{11}$ 
& $2.17\times10^{3}$ 
& $4.48\times10^{12}$ 
\\
$\delta=10^{-10}$ 
& $2.261\times10^{-6}$ 
& $1.487\times10^{5}$ 
& $1.051\times10^{10}$ 
& $6.85\times10^{1}$ 
& $1.42\times10^{11}$ 
\\
\bottomrule
\end{tabular}
\caption{
For each ultra-relativistic deficit $\delta=1-v_{\rm term}$ we evolve the
$SU(3)_c$ bubble wall using Eq.~\eqref{eq:alpha-spherical-su3c} until
$\tau_{\rm term}=\sigma/\eta$ is reached. The total yields
$N_{\rm tot}^{\rm(int)}$ are shown separately for one representative scalar
degree of freedom and for the four massive gluons, including only their two
transverse polarizations. The scalar calculation uses
$(\omega_+,\omega_-)=(M_s,m_{sf})$, while the massive-gluon calculation uses
$(\omega_+,\omega_-)=(M_g,0)$. Only the $k=0$ mode quanta are included;
higher-$k$ modes would increase the total particle production.
}
\label{tab:su3c-yields-gluons}
\end{table*}
In this subsection we deliberately adopt a simplified description in which the
drag coefficient~$\eta$ is taken to be constant, so that the dimensionless
friction ratio $B=\eta/\sigma$ in Eq.~\eqref{eq:alpha-spherical-su3c} does
not depend on the state of the shocked plasma.  The purpose of this discussion
is to isolate the pure ``vacuum--mismatch'' contribution to particle
production: the accelerating wall imposes a time–dependent boundary condition
on the relevant quantum fields, and the non–adiabatic change in the mode frequencies
$\omega_\pm$ produces quanta even in the absence of an explicitly thermalized
medium.  The resulting zero–mode occupancy $N_{k=0}(\tau)$ in
Eqs.~\eqref{eq:Nk0-alpha-su3c}–\eqref{eq:Nk0-SU3} should therefore be viewed
as a minimal, non–thermal lower bound on the particle yield associated with
the acceleration history of the wall.

Using the benchmark values from Table~\ref{tab:su3c-yields-gluons}, the typical vacuum energy difference is \(\Delta V\simeq0.16~\mathrm{TeV}^{4}\). The total energy released as the wall expands to its final radius is then
\begin{equation}
E_{\text{vac}}
=\Delta V\int_{R_0}^{R_{\text{fin}}} 4\pi R^2\,dR
=\frac{4\pi}{3}\Delta V\big(R_{\text{fin}}^3-R_0^3\big).
\end{equation}
For \(R_{\text{fin}}\gg R_0\), this simplifies to
\begin{equation}
E_{\text{vac}}\simeq \frac{4\pi}{3}\,\Delta V\,R_{\text{fin}}^3,
\end{equation}
numerically
\[
E_{\text{vac}}\approx
\begin{cases}
6.7\times10^{35}\,\mathrm{TeV}, & \delta=10^{-12},\\
6.7\times10^{32}\,\mathrm{TeV}, & \delta=10^{-11},\\
6.7\times10^{29}\,\mathrm{TeV}, & \delta=10^{-10}.
\end{cases}
\]
Even if only a minute fraction of this energy converts into rest mass, it readily accounts for the creation of multi--TeV particles. The accelerating wall imposes a time-dependent boundary condition on the surrounding quantum fields. As this boundary moves, the vacuum configuration changes, and the inertial observer perceives particle creation. Only fields that couple to the boundary respond to this varying condition; uncoupled fields remain unaffected. This process resembles a dynamical Casimir effect, where the moving boundary extracts energy from the vacuum and converts it into real particles. This viewpoint is not in conflict with the earlier Unruh-like interpretation: the Unruh description highlights the role of the wall's proper acceleration, whereas the dynamical-Casimir description highlights the non-adiabatic change of the field modes induced by the moving wall. In the present setup these should be understood as complementary interpretations of the same underlying particle-production mechanism. Thus, the large vacuum pressure driving the bubble expansion provides the true energy reservoir for producing massive quanta in the true-vacuum phase.

As the radius increases, drag compensates the vacuum drive and the wall approaches terminal speed on the timescale $\tau_{\rm term}$. The total yield reflects the competition between the exponential sensitivity of $N_{k=0}(\alpha)$ to the declining acceleration and the rapidly growing area factor $4\pi R^2$, which becomes significant by $\tau\sim\tau_{\rm term}$ in the most ultra–relativistic cases.
\section{Thermal particle production from frictional dissipation in the \texorpdfstring{$SU(3)_c$}{SU(3)c} transition}
\label{sec:thermal-su3}

As the $SU(3)_c$-breaking bubble wall propagates through an ambient medium, microscopic scatterings between the wall and the surrounding plasma exert a frictional pressure
\begin{equation}
P_{\rm fric} = \eta\,\gamma v,
\end{equation}
which acts against the vacuum--mismatch force responsible for accelerating the wall.  
In the absence of friction the wall would acquire a large kinetic energy; with friction, part of this energy is continuously drained and deposited into a thin layer of shocked plasma immediately behind the wall.  
The purpose of this section is to estimate how this dissipated energy can be converted into additional particles. However, before using a thermal description, one must check whether the corresponding temperature is compatible with the finite-temperature phase diagram of the $SU(3)_c$-breaking model.

This point is important because, for the benchmark considered in this paper, a plasma hot enough to make the massive broken-phase gluons and scalar octet ultrarelativistic would lie above the symmetry-restoration temperature. Therefore the thermal channel cannot be treated as an equilibrated relativistic gas of broken-phase massive gluons and scalar octets. Instead, a finite-temperature-consistent thermal estimate must be restricted to the subcritical regime $T<T_c$, where the broken $SU(3)_c\to U(2)$ phase remains locally stable. In this regime the massive modes can still be produced, but they are nonrelativistic and their abundance is exponentially Boltzmann suppressed.

The treatment presented here follows the same energy--deficit logic as in our previous false Higgs vacuum decay analysis in Section~7 of Ref.~\cite{Sengupta:2025jah}, but is now supplemented by the finite-temperature effective potential of the adjoint $SU(3)_c$ scalar model~\cite{Stojkovic:2007dw,PhysRevD.19.1906}. We also discuss a separate nonequilibrium wall--matter transition-radiation channel, which does not rely on a thermal bath with $T>T_c$ and can continue after the wall reaches terminal velocity.

\subsection{Finite-temperature restoration and phase diagram}
\label{subsec:su3-finiteT}

The zero-temperature potential used in Sec.~\ref{Model} is
\begin{equation}
V(\psi)
=
\frac94\lambda_1\psi^2(\psi-\psi_0)^2
-
\frac94\epsilon_0\psi_0^2\psi^2
+V_0.
\end{equation}
At one loop, the finite-temperature effective potential can be written in the form~\cite{Stojkovic:2007dw,PhysRevD.19.1906}
\begin{equation}
V_{\rm eff}(\psi,T)
=
\frac94\lambda_1\psi^2(\psi-\psi_0)^2
-
\frac94\epsilon(T)\psi_0^2\psi^2
+V_0,
\label{eq:Veff-finiteT-su3}
\end{equation}
where
\begin{equation}
\epsilon(T)
=
\epsilon_0
-
\left(
\frac59\lambda_1+\frac12 g_s^2
\right)
\frac{T^2}{\psi_0^2}.
\label{eq:epsilonT-su3}
\end{equation}
The coefficients in Eq.~\eqref{eq:epsilonT-su3} are written in the same canonically normalized one-field convention as Eq.~\eqref{eq:canonical-reduced-potential}. The critical temperature is defined by $\epsilon(T_c)=0$, giving
\begin{equation}
T_c^2
=
\frac{2\epsilon_0\psi_0^2}{10\lambda_1/9+g_s^2}.
\label{eq:Tc-su3}
\end{equation}
For the benchmark values used throughout this paper,
\begin{equation}
\epsilon_0=0.076,\qquad
\lambda_1=1,\qquad
\psi_0=0.983271~{\rm TeV},
\end{equation}
and with the QCD coupling evaluated at the TeV scale,
\begin{equation}
\alpha_s(1~{\rm TeV})\simeq0.09,
\qquad
g_s=\sqrt{4\pi\alpha_s}\simeq1.06,
\end{equation}
Eq.~\eqref{eq:Tc-su3} gives
\begin{equation}
T_c
\simeq
0.256~{\rm TeV}.
\label{eq:Tc-numeric-su3}
\end{equation}

This should be compared with the broken-phase masses
\begin{equation}
M_g\simeq1.0~{\rm TeV},
\qquad
M_s\simeq2.48~{\rm TeV}.
\end{equation}
Thus,
\begin{equation}
T_c < M_g < M_s.
\label{eq:Tc-mass-hierarchy-su3}
\end{equation}
Consequently, an equilibrated thermal bath hot enough to make the massive gluons or scalar octets ultrarelativistic would lie in the symmetry-restored regime.

The disappearance of the broken minimum occurs at a somewhat higher temperature. Defining
\begin{equation}
x\equiv\frac{\psi}{\psi_0},
\end{equation}
the nonzero extrema of Eq.~\eqref{eq:Veff-finiteT-su3} satisfy
\begin{equation}
\lambda_1(2x^2-3x+1)-\epsilon(T)=0.
\end{equation}
A nonzero extremum exists only if
\begin{equation}
1+\frac{8\epsilon(T)}{\lambda_1}\ge 0.
\end{equation}
This defines the spinodal, or superheating, temperature
\begin{equation}
T_{\rm sp}^2
=
\frac{2(\epsilon_0+\lambda_1/8)\psi_0^2}
{10\lambda_1/9+g_s^2}.
\label{eq:Tsp-su3}
\end{equation}
For the same benchmark,
\begin{equation}
T_{\rm sp}
\simeq
0.416~{\rm TeV}.
\label{eq:Tsp-numeric-su3}
\end{equation}

The finite-temperature phase structure is therefore
\begin{align}
T<T_c:
&\qquad
SU(3)_c\to U(2)\ \text{broken phase is thermodynamically favored},
\nonumber\\[1ex]
T_c<T<T_{\rm sp}:
&\qquad
\text{the broken phase can exist only as a superheated metastable phase},
\nonumber\\[1ex]
T>T_{\rm sp}:
&\qquad
\text{the nonzero broken minimum disappears and }SU(3)_c\text{ is restored}.
\label{eq:phase-diagram-su3}
\end{align}

This phase diagram fixes the correct interpretation of thermal particle production. For $T<T_c$, massive broken-phase gluons and scalar octets are good quasiparticles, but they are nonrelativistic because $T_c<M_g,M_s$. They can be thermally produced only through the Boltzmann tail. For $T>T_c$, an equilibrated plasma is no longer described by broken-phase massive gluons and scalar octets. In the restored phase, the relevant degrees of freedom are those of the color-symmetric phase, and the decay phenomenology of the massive broken-phase states no longer applies.

\subsection*{Frictional heating, shocked-layer temperature, and the role of the efficiency factor}

The friction term in the wall equation of motion represents energy and
momentum transfer from the accelerating wall to the surrounding medium. In
the thin-wall description the wall energy per unit area is
\begin{equation}
E_{\rm wall}(t)=\sigma\gamma(t),
\qquad
\gamma(t)=\frac{1}{\sqrt{1-v^2(t)}}.
\end{equation}
If the wall were frictionless, the vacuum pressure would continue to increase
the wall Lorentz factor. With friction, part of this would-be kinetic energy
is instead deposited into a shocked layer near the wall. This dissipated
energy is the physical reservoir from which additional particles may be
produced.

A useful order-of-magnitude estimate of the characteristic temperature of this layer can be
obtained by assuming that a fraction \(\kappa_{\rm sh}\le1\) of the released
vacuum energy density is temporarily stored as radiation in the shocked region,
\begin{equation}
\rho_{\rm sh}
\simeq
\kappa_{\rm sh}\Delta V
=
\frac{\pi^2}{30}g_*T_{\rm sh}^4,
\end{equation}
where \(g_*\) is the effective number of relativistic degrees of freedom in
the shocked layer. This gives
\begin{equation}
T_{\rm sh}
\simeq
\left(
\frac{30\,\kappa_{\rm sh}\Delta V}{\pi^2 g_*}
\right)^{1/4}.
\label{eq:Tshock-estimate}
\end{equation}
For the benchmark value \(\Delta V\simeq0.16~{\rm TeV}^4\), and for the
reference choice \(g_*=100\), this becomes
\begin{equation}
T_{\rm sh}
\simeq
0.264~{\rm TeV}\,
\kappa_{\rm sh}^{1/4}
\left(\frac{100}{g_*}\right)^{1/4}.
\label{eq:Tshock-numeric}
\end{equation}
For the reference choice \(\kappa_{\rm sh}=1\) and \(g_*=100\), this value is very close to the critical temperature
\begin{equation}
T_c\simeq0.256~{\rm TeV}.
\end{equation}
Thus the shocked layer is in a marginal regime: order-one changes in
\(\kappa_{\rm sh}\), \(g_*\), the shell lifetime, and the hydrodynamic cooling
can decide whether the layer is temporarily above or below the color-restoration
temperature.

If \(T_{\rm sh}>T_c\), the local broken minimum is not thermodynamically
favored. In that case the layer should not be described as an equilibrated gas
of broken-phase massive gluons and color-octet scalars. The relevant degrees
of freedom are instead those of the restored color-symmetric phase. Massive
broken-phase particles can only be counted after the layer cools back below
\(T_c\), where their abundance is controlled by the subcritical Boltzmann
tail. Thus, in this case our thermal estimate should be read as a freeze-out
estimate evaluated at \(T_*<T_c\), not as a high-temperature equilibrium yield.

If \(T_{\rm sh}<T_c\), the broken \(SU(3)_c\to U(2)\) phase remains locally
stable. Massive broken-phase gluons and color-octet scalars are then meaningful
quasiparticles, but they are still nonrelativistic because
\begin{equation}
T_{\rm sh}<T_c<M_g<M_s.
\end{equation}
Their thermal abundance is therefore Boltzmann suppressed. In the numerical
estimate below we take
\begin{equation}
T_*=0.9T_c=0.230~{\rm TeV}
\end{equation}
as a representative subcritical freeze-out temperature.

There is also a finite-width limitation. The shocked layer is thin, and if its width is of order the Compton wavelength of the heavy particle being produced or smaller, then it cannot be treated as a macroscopic equilibrium heat bath. For a particle
species \(i\) with mass \(m_i\),
\begin{equation}
\lambda_{C,i}\sim\frac{1}{m_i},
\end{equation}
while the nonrelativistic thermal wavelength at \(T_*\) is
\begin{equation}
\lambda_{{\rm th},i}
\sim
\sqrt{\frac{2\pi}{m_iT_*}}.
\end{equation}
For \(T_*=0.230~{\rm TeV}\), the benchmark values are
\begin{equation}
\lambda_{C,g}\simeq1.0~{\rm TeV}^{-1},
\qquad
\lambda_{{\rm th},g}\simeq5.2~{\rm TeV}^{-1},
\end{equation}
for the massive gluon, and
\begin{equation}
\lambda_{C,G_H}\simeq0.40~{\rm TeV}^{-1},
\qquad
\lambda_{{\rm th},G_H}\simeq3.3~{\rm TeV}^{-1},
\end{equation}
for the color-octet scalar. Therefore, if the shocked layer has thickness
\begin{equation}
\ell_{\rm sh}\sim \lambda_{C,i},
\end{equation}
then the layer is thinner than the corresponding thermal formation length. A
simple way to parametrize this suppression is
\begin{equation}
f_{{\rm width},i}
\sim
\min\left(1,\frac{\ell_{\rm sh}}{\lambda_{{\rm th},i}}\right).
\end{equation}
For a Compton-scale layer this gives roughly
\begin{equation}
f_{{\rm width},g}\sim0.2,
\qquad
f_{{\rm width},G_H}\sim0.1.
\end{equation}
Moreover, a perturbative estimate of the mean free path at \(T_*\) gives
\begin{equation}
\ell_{\rm mfp}
\sim
\frac{1}{\alpha_s^2T_*}
\sim
5\times10^2~{\rm TeV}^{-1},
\end{equation}
which is much larger than a Compton-scale shocked layer. This indicates that
full kinetic equilibration of the heavy particles inside such a thin layer is
not expected.

For these reasons, the thermal contribution below should be understood as a
finite-temperature-consistent freeze-out estimate, not as the solution of a
fully dynamical high-temperature plasma problem. We absorb incomplete
thermalization, finite shell lifetime, hydrodynamic dilution, and the finite
width of the shocked layer into the efficiency factor
\begin{equation}
f_{\rm th}
=
f_{\rm heat}\,f_{\rm width}\,f_{\rm eq}
\le1.
\end{equation}
For compactness we denote these efficiencies collectively by \(f_{\rm th}\), although in a microscopic treatment they can differ between the scalar and massive-gluon channels. For a Compton-scale shocked layer one expects \(f_{\rm th}\) to be substantially below unity. The numerical yields below are therefore quoted per unit \(f_{\rm th}\), so that they can be rescaled once the microscopic shock structure is known. 

\subsection{Subcritical thermal production of massive broken-phase quanta}
\label{subsec:subcritical-thermal-su3}

Thermal production of massive gluons and color-octet scalars in the broken phase is allowed only while the local temperature satisfies
\begin{equation}
T<T_c.
\end{equation}
In this regime the particles are not ultrarelativistic. Their equilibrium number density is therefore not the relativistic expression proportional to $T^3$, but the massive Bose--Einstein density
\begin{equation}
n_i(T)
=
\frac{d_i\,m_i^2T}{2\pi^2}
\sum_{\ell=1}^{\infty}
\frac{1}{\ell}
K_2\!\left(\ell\frac{m_i}{T}\right),
\qquad
i=s,g,
\label{eq:massive-BE-density-su3}
\end{equation}
where $K_2$ is the modified Bessel function. Since $m_i/T>1$ throughout the subcritical regime, the leading Bessel term is an excellent approximation,
\begin{equation}
n_i(T)
\simeq
\frac{d_i\,m_i^2T}{2\pi^2}
K_2\!\left(\frac{m_i}{T}\right),
\end{equation}
while the asymptotic Maxwell--Boltzmann form gives the useful analytic scaling
\begin{equation}
n_i(T)
\simeq
d_i
\left(
\frac{m_iT}{2\pi}
\right)^{3/2}
e^{-m_i/T}.
\label{eq:MB-density-su3}
\end{equation}
We use
\begin{equation}
M_g=1.0~{\rm TeV},
\qquad
d_g=8,
\end{equation}
for the four massive gluons with two transverse polarizations each. This is a conservative choice consistent with the transverse-only treatment used in the vacuum-mismatch calculation; including the longitudinal polarizations of a fully thermalized massive vector would increase the massive-gluon thermal yield by a factor $12/8=3/2$. For one representative physical scalar mode we use
\begin{equation}
M_s=2.48~{\rm TeV},
\qquad
d_s=1.
\end{equation}

For definiteness, we take the subcritical freeze-out temperature to be
\begin{equation}
T_*=\xi T_c,
\qquad
\xi=0.9,
\end{equation}
so that
\begin{equation}
T_*
=
0.230~{\rm TeV}
<
T_c.
\label{eq:Tstar-su3}
\end{equation}
At this temperature,
\begin{equation}
\frac{M_g}{T_*}=4.34,
\qquad
\frac{M_s}{T_*}=10.77.
\end{equation}
Thus the massive gluons are Boltzmann suppressed by roughly $e^{-4.34}$, while the scalar is much more strongly suppressed by roughly $e^{-10.77}$.

Using Eq.~\eqref{eq:massive-BE-density-su3}, one finds
\begin{equation}
n_g(T_*)
\simeq
1.08\times10^{-3}~{\rm TeV}^3,
\qquad
n_s(T_*)
\simeq
6.83\times10^{-7}~{\rm TeV}^3.
\label{eq:ngns-subcritical}
\end{equation}
The scalar density quoted here is for one physical scalar degree of freedom. If the four physical scalars are approximately degenerate, the total scalar yield should be multiplied by four.

As the wall advances by an amount \(dR\), it sweeps out the spherical shell volume
\begin{equation}
dV = 4\pi R^2\,dR .
\end{equation}
Here \(dR\) is the infinitesimal increase in the bubble radius during the corresponding time interval. In the proper-time parametrization,
\begin{equation}
dR=\frac{dR}{d\tau}d\tau=\sinh y(\tau)\,d\tau,
\end{equation}
so the same swept volume can also be written as
\begin{equation}
dV
=
4\pi R^2(\tau)\sinh y(\tau)\,d\tau .
\end{equation}
We parametrize the fraction of the swept region that actually passes through a subcritical thermal freeze-out stage by the efficiency factor \(f_{\rm th}\le1\) discussed above. The subcritical thermal yield is then
\begin{equation}
N_i^{(<T_c)}
=
f_{\rm th}
\int_{R_0}^{R_{\rm final}}
4\pi R^2dR\;n_i(T_*),
\end{equation}
or
\begin{equation}
N_i^{(<T_c)}
=
f_{\rm th}
\frac{4\pi}{3}
\left(
R_{\rm final}^3-R_0^3
\right)
n_i(T_*).
\label{eq:subcritical-yield-su3}
\end{equation}
If one counted only a microscopic shocked layer of
thickness \(\ell_{\rm sh}\) at the final radius, the relevant volume would be
\begin{equation}
V_{\rm shell}\sim 4\pi R_{\rm final}^2\ell_{\rm sh}.
\end{equation}
By contrast, Eq.~\eqref{eq:subcritical-yield-su3} counts the cumulative
volume swept by the wall during its expansion,
\begin{equation}
V_{\rm swept}
=
\frac{4\pi}{3}
\left(R_{\rm final}^3-R_0^3\right).
\end{equation}
For \(R_{\rm final}\gg R_0\), the ratio is approximately
\begin{equation}
\frac{V_{\rm swept}}{V_{\rm shell}}
\sim
\frac{R_{\rm final}}{3\ell_{\rm sh}}.
\end{equation}
Thus the calculation has two competing effects. The local thermal density is reduced by the Boltzmann factors because
\(T_*<T_c<M_g<M_s\), but the cumulative swept volume can be much larger than the volume of a single microscopic shell. The yields quoted below are therefore best interpreted as per-efficiency upper-envelope estimates, with the finite width, finite lifetime, and non-equilibration of the shocked layer
absorbed into \(f_{\rm th}\).

For consistency with the previous frictional benchmark, we take
\begin{equation}
R_{\rm final}
\simeq
\gamma_{\rm term}v_{\rm term}\tau_{\rm final},
\qquad
\tau_{\rm final}=5\tau_{\rm term}.
\end{equation}
This should be interpreted as a finite-temperature-consistent upper envelope for broken-phase thermal production, not as a full hydrodynamic calculation of the shocked shell.

\subsection{Numerical results for subcritical thermal production}

For the $SU(3)_c\to U(2)$ benchmark we use
\begin{equation}
\Delta V = 1.60\times10^{-1}~{\rm TeV}^4,\qquad
\sigma = 3.36\times10^{-1}~{\rm TeV}^3,\qquad
R_0 = 6.31~{\rm TeV}^{-1},
\end{equation}
and the three terminal-velocity deficits
\begin{equation}
\delta=10^{-12},\qquad10^{-11},\qquad10^{-10}.
\end{equation}
For each $\delta$ we use the same $\eta$ and $\tau_{\rm final}=5\tau_{\rm term}$ as in the previous frictional analysis. The resulting subcritical thermal yields are shown in Table~\ref{tab:su3c-thermal-yields}.

\begin{table}[htbp]
\centering
\scriptsize
\setlength{\tabcolsep}{4pt}
\renewcommand{\arraystretch}{1.0}
\begin{tabular}{|lccccc|}
\hline
$\delta$
& $\eta$
& $\tau_{\rm final}$
& $T_*$
& \begin{tabular}[c]{@{}c@{}}Scalar yield\\ $N_s^{(<T_c)}/f_{\rm th}$\end{tabular}
& \begin{tabular}[c]{@{}c@{}}Gluon yield\\ $N_g^{(<T_c)}/f_{\rm th}$\end{tabular}
\\
& $[{\rm TeV}^4]$
& $[{\rm TeV}^{-1}]$
& $[{\rm TeV}]$
& 
& 
\\
\hline
$10^{-12}$ &
$2.26\times10^{-7}$ &
$7.43\times10^{6}$ &
$0.230$ &
$4.16\times10^{32}$ &
$6.59\times10^{35}$
\\
$10^{-11}$ &
$7.15\times10^{-7}$ &
$2.35\times10^{6}$ &
$0.230$ &
$4.16\times10^{29}$ &
$6.59\times10^{32}$
\\
$10^{-10}$ &
$2.26\times10^{-6}$ &
$7.43\times10^{5}$ &
$0.230$ &
$4.16\times10^{26}$ &
$6.59\times10^{29}$
\\
\hline
\end{tabular}
\caption{
Subcritical thermal yields in the color-octet scalar channel and in the massive-gluon channel, computed with $T_*=0.9T_c=0.230~{\rm TeV}<T_c$. The massive number densities are evaluated using Eq.~\eqref{eq:massive-BE-density-su3}. The leading Maxwell--Boltzmann expression in Eq.~\eqref{eq:MB-density-su3} gives the same exponential scaling and is useful analytically, while the displayed numbers use the Bessel-function density. The factor $f_{\rm th}\le1$ parametrizes the efficiency with which the dissipated wall energy passes through a subcritical broken-phase thermal stage and freezes out into heavy modes. The scalar column is for one representative physical scalar degree of freedom.
}
\label{tab:su3c-thermal-yields}
\end{table}

The numbers in Table~\ref{tab:su3c-thermal-yields} are quoted per unit efficiency \(f_{\rm th}\). They are therefore not direct predictions of a fully thermalized plasma calculation. They show the size of the possible broken-phase freeze-out yield if a fraction \(f_{\rm th}\) of the swept region passes through a subcritical thermal stage with \(T_*=0.9T_c\). The actual value of \(f_{\rm th}\) requires a dedicated nonequilibrium hydrodynamic treatment. The massive gluon yield remains much larger than the scalar yield because $M_g<M_s$, so the Boltzmann suppression is milder, and because the massive-gluon degeneracy is larger. For \(f_{\rm th}\) not too small, this subcritical thermal channel can still exceed the direct vacuum-mismatch yield in the benchmark estimates, while remaining compatible with the finite-temperature phase diagram. However, if the shocked layer is only of order the Compton wavelength of the heavy particles, the finite-width and non-equilibration effects discussed above imply \(f_{\rm th}\ll1\). In that case the subcritical thermal channel should be regarded as an upper bound, and the more robust dominant source for the benchmark considered here is the non-equilibrium wall--matter transition-radiation channel.

\section{Particle production from non-equilibrium wall--matter transition radiation}
\label{sec:transition-radiation-su3}

The subcritical thermal estimate above is not the only way to obtain additional particles beyond vacuum mismatch. A separate source arises from ambient matter crossing the moving bubble wall. This process is intrinsically nonequilibrium and should be distinguished from a thermalized shocked layer.

In the wall rest frame, even cold ambient matter appears as a highly boosted beam. For a wall velocity
\begin{equation}
v_w=1-\delta,
\end{equation}
the wall Lorentz factor is
\begin{equation}
\gamma_w
=
\frac{1}{\sqrt{1-v_w^2}}
\simeq
\frac{1}{\sqrt{2\delta}}.
\label{eq:gammaw-delta}
\end{equation}
A baryon of mass $m_N$ at rest in the ambient medium therefore has wall-frame energy
\begin{equation}
E_N'\simeq \gamma_w m_N.
\end{equation}
At the parton level, a valence quark carries a fraction $x_q$ of this energy,
\begin{equation}
E_q'
\simeq
x_q\gamma_w m_N.
\label{eq:Eqprime-su3}
\end{equation}
For the estimates below we take
\begin{equation}
m_N=0.938~{\rm GeV},
\qquad
x_q=\frac13.
\end{equation}
Here $m_N$ is the nucleon mass and $x_q=1/3$ is a simple valence-quark benchmark in which the nucleon energy is shared equally among three valence quarks. A more realistic treatment would fold the transition-radiation probability with parton distribution functions evaluated at a scale set by the wall thickness or by the emitted heavy-particle mass.

When such a quark crosses the $SU(3)_c\to U(2)$ wall, its color-field configuration changes because four gluons acquire mass in the broken phase. The wall is static in its own rest frame, so energy is conserved, but translation invariance in the wall-normal direction is broken. The wall can therefore absorb the longitudinal momentum mismatch required for wall-assisted radiation. This is the color analogue of transition radiation from a charged particle crossing an interface, and of transition radiation from particles crossing relativistic bubble walls in first-order phase transitions~\cite{Moore:1995ua,Moore_2000,Bodeker:2017cim,Hoche:2020ysm,Azatov:2023xem,Ai:2023see}.

The leading process for massive gluon emission is schematically
\begin{equation}
q+\text{wall}\to q+G_{\rm massive}+\text{wall}.
\end{equation}
A first-principles calculation of this process would require solving the
quark and gauge-field scattering problem in the smooth
\(SU(3)_c\to U(2)\) wall background. Here we use a leading-log estimate
motivated by ordinary soft/collinear gauge-boson radiation and by transition
radiation from particles crossing relativistic bubble walls. The emission
probability is written as
\begin{equation}
P_{q\to qG}
\simeq
f_G\,
\frac{C_F\alpha_s}{\pi}
\ln\!\left(\frac{E_q'^2}{M_g^2}\right)
\Theta(E_q'-M_g),
\label{eq:PqG-transition}
\end{equation}
where
\begin{equation}
C_F=\frac43,
\qquad
\alpha_s=0.09.
\end{equation}
Here \(\Theta(E_q'-M_g)\) is the Heaviside step function. It enforces the
kinematic threshold for real massive-gluon emission in the wall frame:
\begin{equation}
\Theta(E_q'-M_g)
=
\begin{cases}
1, & E_q'>M_g,\\
0, & E_q'<M_g.
\end{cases}
\end{equation}
The wall can absorb momentum perpendicular to the interface because translational
invariance in that direction is broken, but the wall is static in its own rest
frame and therefore energy is conserved. Hence an incoming quark with wall-frame
energy below \(M_g\) cannot radiate an on-shell massive gluon. The step function
also prevents the leading-log expression from being used below threshold, where
the logarithm would no longer describe a physical real-emission probability.

The factor \(f_G\le1\) parametrizes finite wall thickness, color projection,
hadronic coherence, and other model-dependent effects. This expression should
therefore be interpreted as an order-of-magnitude benchmark rather than an exact
microscopic prediction. The soft/collinear QCD structure and the color factor
\(C_F=4/3\) are standard~\cite{ParticleDataGroup:2022pth}, while the use of
transition radiation from particles crossing relativistic bubble walls is
motivated by Refs.~\cite{Bodeker:2017cim,Hoche:2020ysm,Ai:2023see}. For
\(f_G\sim1\), the leading-log probability can become order unity for the most
relativistic benchmark. In that regime the single-emission estimate should be
interpreted as an inclusive order-of-magnitude rate, and a more complete
treatment would resum multiple emissions and enforce unitarity. In the present
work this uncertainty is absorbed into the efficiency factor \(f_G\).

The origin of Eq.~\eqref{eq:PqG-transition} is the usual leading-log
soft/collinear radiation estimate, adapted to the wall-crossing problem.
In the wall frame, the incoming quark is a high-energy color charge crossing
an interface where part of the gauge spectrum becomes massive. In the region
where the emitted massive gluon is approximately collinear with the incoming
quark, the radiation probability factorizes from the hard wall-scattering
process and has the schematic form
\begin{equation}
dP_{q\to qG}
\simeq
f_G\,
\frac{C_F\alpha_s}{\pi}\,
\frac{dk_\perp^2}{k_\perp^2+M_g^2}.
\end{equation}
The mass \(M_g\) cuts off the collinear singularity in the infrared, while
the wall-frame quark energy \(E_q'\) sets the ultraviolet scale of the
transverse momentum integral. Therefore, above threshold,
\begin{equation}
P_{q\to qG}
\sim
f_G\,
\frac{C_F\alpha_s}{\pi}
\int_{M_g^2}^{E_q'^2}
\frac{dk_\perp^2}{k_\perp^2}
=
f_G\,
\frac{C_F\alpha_s}{\pi}
\ln\!\left(\frac{E_q'^2}{M_g^2}\right).
\end{equation}
Including the threshold condition gives Eq.~\eqref{eq:PqG-transition}. A
calculation with the full splitting kernel, a finite wall profile, and hadronic
wave functions would replace the sharp step function by a smoother threshold
behavior and modify the order-one prefactor; these effects are included in the
phenomenological efficiency factor \(f_G\).

For the physical color-octet scalar, the corresponding process is
\begin{equation}
q+\text{wall}\to q+G_H+\text{wall}.
\end{equation}
Using the same effective Yukawa normalization as in the phenomenological scalar analysis below,
\begin{equation}
y_q=\frac{m_q}{v},
\qquad
v=246~{\rm GeV},
\end{equation}
we estimate
\begin{equation}
P_{q\to qG_H}
\simeq
f_H\,
\frac{y_q^2}{16\pi^2}
\ln\!\left(\frac{E_q'^2}{M_s^2}\right)
\Theta(E_q'-M_s),
\label{eq:PqH-transition}
\end{equation}
where $f_H\le1$ is the corresponding scalar-emission efficiency factor.

The number of ambient baryons swept up by a bubble expanding to radius $D$ is
\begin{equation}
N_b
=
n_b\frac{4\pi}{3}D^3.
\label{eq:Nb-swept}
\end{equation}
For the reference values
\begin{equation}
n_b=0.25~{\rm m}^{-3},
\qquad
D=10^9~{\rm ly},
\end{equation}
we take \(n_b\) to be the mean baryon number density of the universe today, estimated from
\begin{equation}
n_{b,0}\simeq \frac{\Omega_b\rho_c}{m_p}
=
\frac{3H_0^2\Omega_b}{8\pi G m_p}.
\end{equation}
The distance \(D=10^9~{\rm ly}\) is used as a fiducial cosmological propagation scale. Since the total number of baryons swept up by the wall scales as \(N_b\propto n_bD^3\), results for other distances can be obtained by the simple rescaling \((D/10^9~{\rm ly})^3\). For the reference values above, one finds
\begin{equation}
N_b=8.87\times10^{74}
\left(\frac{n_b}{0.25~{\rm m}^{-3}}\right)
\left(\frac{D}{10^9~{\rm ly}}\right)^3.
\label{eq:Nb-numeric}
\end{equation}
For proton-like matter, with valence content $uud$, the massive-gluon yield is
\begin{equation}
N_G^{\rm rad}
\simeq
3N_b\,P_{q\to qG},
\label{eq:NG-rad}
\end{equation}
while the scalar yield is
\begin{equation}
N_{s}^{\rm rad}
\simeq
N_b\left(2P_{u\to u {G_H}}+P_{d\to d {G_H}}\right).
\label{eq:NH-rad}
\end{equation}
The massive-gluon channel is much less suppressed because it is controlled by the gauge coupling rather than by light-quark Yukawa couplings.

For the three terminal-velocity deficits used elsewhere in the paper, Eq.~\eqref{eq:gammaw-delta} gives
\begin{equation}
\gamma_w=
7.07\times10^5,\quad
2.24\times10^5,\quad
7.07\times10^4,
\end{equation}
for
\begin{equation}
\delta=10^{-12},\quad10^{-11},\quad10^{-10},
\end{equation}
respectively. The corresponding wall-frame valence-quark energies are
\begin{equation}
E_q'=221~{\rm TeV},\quad69.9~{\rm TeV},\quad22.1~{\rm TeV}.
\end{equation}
These energies are above both the massive-gluon threshold $M_g\simeq1~{\rm TeV}$ and the scalar threshold $M_s\simeq2.48~{\rm TeV}$.

Using
\begin{equation}
m_u=2.2~{\rm MeV},
\qquad
m_d=4.7~{\rm MeV},
\end{equation}
we obtain the transition-radiation yields shown in Table~\ref{tab:transition-radiation-yields}.

\begin{table}[t]
\centering
\small
\begin{tabular}{@{}lcccc@{}}
\toprule
Scenario ($\delta$)
& $\gamma_w$
& $E_q'$ [TeV]
& $N_G^{\rm rad}/f_G$
& $N_s^{\rm rad}/f_H$
\\
\midrule
$10^{-12}$ &
$7.07\times10^5$ &
$221$ &
$1.10\times10^{75}$ &
$2.65\times10^{64}$
\\
$10^{-11}$ &
$2.24\times10^5$ &
$69.9$ &
$8.63\times10^{74}$ &
$1.97\times10^{64}$
\\
$10^{-10}$ &
$7.07\times10^4$ &
$22.1$ &
$6.29\times10^{74}$ &
$1.29\times10^{64}$
\\
\bottomrule
\end{tabular}
\caption{
Order-of-magnitude wall--matter transition-radiation yields for a bubble expanding through an ambient baryon density $n_b=0.25~{\rm m}^{-3}$ over a distance $D=10^9~{\rm ly}$. The massive-gluon estimate uses Eq.~\eqref{eq:PqG-transition}, while the scalar estimate uses Eq.~\eqref{eq:PqH-transition}. The unknown efficiency factors $f_G$ and $f_H$ encode wall-profile, color-projection, finite-thickness, and hadronic-coherence effects. The scalar column assumes proton-like valence content $uud$.
}
\label{tab:transition-radiation-yields}
\end{table}

This channel is conceptually different from the thermal channel. The large energy scale is the wall-frame beam energy $E_q'$, not a plasma temperature. Therefore it does not require a thermal bath with $T>T_c$ and is not excluded by the finite-temperature restoration argument. It can also continue during the terminal-velocity stage, since it only requires ambient matter to cross the wall, not a nonzero wall acceleration.

The estimates in Table~\ref{tab:transition-radiation-yields} should not be interpreted as a first-principles prediction. They demonstrate that, because a cosmological wall can sweep through an enormous number of baryons, even an extremely small transition-radiation efficiency can dominate over the direct vacuum-mismatch yield. A full microscopic calculation would require solving the confined hadronic scattering problem across the $SU(3)_c\to U(2)$ wall and matching the corresponding color and gauge-field modes. The leading-log calculation above provides an order-of-magnitude benchmark for this nonequilibrium source.

The photon and neutrino spectra from transition-radiation production are obtained by multiplying the per-parent \texttt{Pythia} spectra by the parent multiplicities,
\begin{equation}
\frac{dN_{\gamma,\nu}^{\rm rad}}{dE}
=
N_G^{\rm rad}
\frac{dN_{\gamma,\nu}^{(g)}}{dE}
+
N_s^{\rm rad}
\frac{dN_{\gamma,\nu}^{(s)}}{dE}.
\label{eq:transition-radiation-spectrum}
\end{equation}
Since the transition-radiation calculation presented here gives only the total parent multiplicity and not the full parent energy distribution, the spectra shown below should be interpreted as rest-frame decay and hadronization spectra normalized to the transition-radiation yield. A complete calculation of the source-frame photon and neutrino spectrum would require folding the \texttt{Pythia} decay spectra with the energy distribution of the radiated massive gluons and scalar octets. Unlike the subcritical thermal contribution, this signal is explicitly environment-dependent:
\begin{equation}
N^{\rm rad}
\propto
n_bD^3.
\end{equation}
It is therefore strongest when the bubble wall propagates through dense matter, and weakest in cosmic voids.

\section{Overview of TeV‐Scale color states}

In addition to the colored scalar field, $\Phi$, once the symmetry is broken to $U(2)$ there emerge both massive and massless gluons. The four massive gluons behave analogously to the $W$ bosons in $SU(2)$: they acquire mass through the breaking and do not form glueballs. However, they can form bound states through the residual gauge force mediated by the unbroken gauge bosons, much like the $W^+$ and $W^-$ in the Standard Model can form a bound state via their mutual Coulomb interaction. These massive–gluon bound states are expected to have masses in the multi‐TeV range—approximately the sum of the constituent masses—and sizes set by the infrared scale of the unbroken non-Abelian subgroup, analogous to charmonium ($c\bar{c}$) or bottomonium ($b\bar{b}$) systems. 

Of the four massless gauge bosons of the unbroken \(U(2)\), the three associated with the non-Abelian \(SU(2)\subset U(2)\) are expected to participate in the infrared confining dynamics and can form glueball-like bound states. The remaining Abelian \(U(1)\) gauge boson instead mediates a long-range Coulombic
interaction. We do not further analyze the decay of the massive–gluon bound state, since its final‐state signatures would be indistinguishable from those of the massive gluon decays. Mixed color-singlet bound states containing one adjoint scalar require at least a quark--antiquark pair, since \(8\otimes 3\) does not contain a singlet whereas \(8\otimes3\otimes\bar3\) does. Such scalar--\(q\bar q\) bound states can carry integer electric charge, while scalars in higher \(SU(3)_c\) representations \((3,6,10,15,21)\) can even form exotic fractionally charged hadrons. Although our analysis is primarily cosmological, we note that the collider interpretation of these states is model-dependent, since the broken-phase gauge spectrum and couplings differ from ordinary color-octet resonance models. Existing ATLAS and CMS searches for colored resonances typically constrain masses in the $\mathcal{O}(1)$ TeV range, with the precise bound depending on branching fractions, widths, and production channels. Our scalar benchmark at $M_s\simeq2.5$ TeV lies above the quoted range, while the $M_g\simeq1$ TeV massive-gluon benchmark should be understood as an illustrative cosmological benchmark rather than a dedicated collider-safe point. A full collider reinterpretation is beyond the scope of this work. In the following sections, we will detail the mass spectra, decay widths, and observational signatures of these TeV‐scale colored states~\cite{Vignaroli:2011ik,Vignaroli:2011um,Vignaroli:2015ama,Bini:2011zb,Luna:2010zza,Barcelo:2011wu}.

All spectra were generated with \texttt{Pythia 8 (v8.313)}\cite{Bierlich:2022pfr,Sjostrand:2014zea} using a fixed random seed and the Monash $2013$ tune. To isolate the intrinsic particle–physics yields from model parameters and avoid dependence on collider or PDF assumptions, the new states (massive gluon, color-octet scalar) were generated as resonances at rest with their target mass, fixed width, and benchmark branching fractions. Parton showering, hadronization, and all weak/FSR decays were enabled, and long-lived mesons and muons were forced to decay so that neutrino and photon yields are fully captured. Each spectrum was produced with $\mathcal{O}(10^4)$ simulated decays, histogrammed in $1$~GeV bins up to several TeV, and written to CSV. This procedure ensures that the photon and neutrino spectra shown in the figures represent the intrinsic rest-frame decay yields of the TeV-scale colored states, independent of collider environments, while still incorporating the complete hadronization and decay chains.

Inside the bubble, in the new vacuum where the symmetry is broken, the fields need not form ordinary \(SU(3)_c\)-singlet bound states. Their long-distance behavior is instead governed by the residual \(U(2)\) gauge dynamics; in particular, states charged under the unbroken \(SU(2)\) sector may still
participate in residual confinement. However, outside of the bubble, in the false vacuum the symmetry is still unbroken, so colored partons produced by the decay of the heavy states cannot remain free and must ultimately hadronize into SU(3)$_c$ singlet final states. In practice, the heavy colored resonance may decay before hadronization, after which its colored decay products shower and hadronize. This is the sequence modeled by \texttt{Pythia}.

When particles are produced at the boundary of an expanding bubble during a first-order phase transition, a subset of them will propagate outward into the false-vacuum region and another fraction will fall into the true-vacuum interior of the bubble. Those particles entering the true vacuum may themselves decay following the symmetries of the true vacuum, and in principle some of their lighter decay products could overtake the advancing bubble wall and reach the false-vacuum side; however, this fraction is extremely suppressed. In the present work, we focus exclusively on the population of particles that decay in the vicinity of the bubble wall and whose decay products propagate outward toward an observer located in the false vacuum. The fraction of inward-going particles that later produce outward-moving decay products will be negligible, and thus we consistently ignore it in our analysis.
\subsection{Phenomenology of the color‐octet scalar \texorpdfstring{$G_H$}{GH}}
\label{subsec:color-octet}
In extensions of QCD with an adjoint scalar acquiring a TeV‐scale VEV that we are considering here, the physical scalar octet $G_H^a$ has mass (see Eq.~\ref{Mass_True_Vacuum})
\begin{equation}
  M_s\;=\;2.5~\mathrm{TeV},
  \qquad \text{for} \qquad
  u\;=\;\langle\Phi\rangle\;=\;1~\mathrm{TeV}\,.
\end{equation}
We work in a traceless diagonal representation, which greatly simplifies the analysis by reducing the dynamics to a single scalar field that plays the role of the order parameter for the symmetry breaking. For this reason, the production formulas presented in the following are written for one representative physical scalar mode. However, in the full adjoint sector of $SU(3)_c$, there are eight real scalar components. After the breaking $SU(3)_c \to U(2)$, four of these are eaten by the four massive gluons, leaving four physical scalar degrees of freedom. Thus, the results shown here should be understood as applying to a single physical scalar mode. If the four physical scalars are approximately degenerate, the total scalar yield is obtained by multiplying the single-mode result by four. If their masses are split, as indicated for example in Eq.~(2.28) of \cite{PhysRevD.19.1906}, the same analysis can be carried out for each mode separately with the appropriate mass inserted.

For the decay rate expressions of the physical color octet scalar we mostly followed \cite{Gresham:2007ri, Gerbush:2007fe, Djouadi:2012ua, Logan:2014jla} and the detailed calculations are shown in Appendix~\ref{App-E} and \ref{App-F}, while for the phenomenological aspects one can refer to \cite{Chivukula:2013hga,SekharChivukula:2014uoe,Chivukula:2015kua,Gerbush:2007fe,Drueke:2014pla,Gresham:2007ri,Dobrescu:2007yp,Simmons:1996fz,Simmons:1996jz,Frampton:1987dn,Davidson:1987mg,FileviezPerez:2008ib,Schumann:2011ji,He:2013tla,Arnold:2011ra,Hayreter:2017wra,Miralles:2019uzg,Idilbi:2010rs,Kim:2008bx,Sayre:2011ed,Hayreter:2018ybt,Krasnikov:1995kn}. In the present broken-color setup, the tree-level decay of a fixed adjoint component $G_H^a$ into quark--antiquark pairs is most cleanly described by introducing an effective Yukawa interaction of the form
\begin{equation}
  \mathcal{L}_{\rm int}\supset -\,y_q\,\bar q\,T^a q\,G_H^a.
\end{equation}
We treat this term as an effective phenomenological interaction that captures the decay of the physical scalar mode into quarks after symmetry breaking. Its microscopic origin is not specified in the present analysis, but it could naturally emerge in a suitable UV-complete extension.

For a single physical color-octet component, the corresponding tree-level partial width is then (see Appendix~\ref{App-E})
\begin{equation}\label{eq:Gammaqq}
  \Gamma\bigl(G_H^a\to q\bar q\bigr)
  =\frac{y_q^2\,M_s}{16\pi}\,
   \Bigl(1-\frac{4m_q^2}{M_s^2}\Bigr)^{3/2}.
\end{equation}
For the numerical benchmark below we adopt the phenomenological normalization
\begin{equation}
  y_q=\eta_q\,\frac{m_q}{v},
  \qquad
  v=246~\mathrm{GeV},
\end{equation}
and set $\eta_q=1$ for all quark flavors. In this normalization, Eq.~\eqref{eq:Gammaqq} becomes
\begin{equation}
  \Gamma\bigl(G_H^a\to q\bar q\bigr)
  =\frac{\eta_q^2\,m_q^2\,M_s}{16\pi v^2}\,
   \Bigl(1-\frac{4m_q^2}{M_s^2}\Bigr)^{3/2}.
\end{equation}
The expression in Eq.~\eqref{eq:Gammaqq} corresponds to the color trace of a fixed octet component, $\Tr(T^aT^a)=T_R=\tfrac12$.\\
For decays into gluons, the leading contribution arises at one loop through quark triangle diagrams. The corresponding partial width for a fixed adjoint component is (see Appendix~\ref{App-F})
\begin{equation}\label{eq:Gamma00}
  \Gamma\bigl(G_H^a\to gg\bigr)
  =\frac{5\,\alpha_s^2\,M_s^3}{6912\,\pi^3}\,
   \left|
     \sum_q \frac{y_q}{m_q}\,F_{1/2}(\tau_q)
   \right|^2,
  \qquad
  \tau_q\equiv \frac{4m_q^2}{M_s^2},
\end{equation}
where $F_{1/2}(\tau_q)$ is the standard spin-$\tfrac12$ scalar loop form factor, given explicitly in Appendix~\ref{App-F}. With the same benchmark choice $y_q=\eta_q\,m_q/v$ and $\eta_q=1$, this may be written as
\begin{equation}
  \Gamma\bigl(G_H^a\to gg\bigr)
  =\frac{5\,\alpha_s^2\,M_s^3}{6912\,\pi^3 v^2}\,
   \left|
     \sum_q \eta_q\,F_{1/2}(\tau_q)
   \right|^2.
\end{equation}
Using 
  $m_u=2.2~\mathrm{MeV},\;
  m_d=4.7~\mathrm{MeV},\;
  m_{s}=96~\mathrm{MeV},\;
  m_c=1.27~\mathrm{GeV},\;
  m_b=4.18~\mathrm{GeV},\;
  m_t=173~\mathrm{GeV}, \text{and} \,
  \alpha_s=0.09,$
together with the benchmark choice $\eta_q=1$, one finds the partial widths and branching ratios shown in Table~\ref{tab:GHdecay}.

\begin{table}[h!]
  \centering
  \begin{tabular}{@{}lcc@{}}
    \toprule
    Channel         & $\Gamma$ [GeV]     & BR (\%)    \\
    \midrule
    $u\bar{u}$      & $3.98\times10^{-9}$ & $<0.01$   \\
    $d\bar{d}$      & $1.82\times10^{-8}$ & $<0.01$   \\
    $s\bar{s}$      & $7.57\times10^{-6}$ & $<0.01$   \\
    $c\bar{c}$      & $1.33\times10^{-3}$ & 0.006     \\
    $b\bar{b}$      & $1.44\times10^{-2}$ & 0.060     \\
    $t\bar{t}$      & $2.39\times10^{1}$  & 99.921    \\
    $gg$            & $3.24\times10^{-3}$ & 0.014     \\
    \midrule
    {\bf Total}     & {\bf $2.39\times10^{1}$} & {\bf 100} \\
    \bottomrule
  \end{tabular}
  \caption{\label{tab:GHdecay}
    Partial widths and branching fractions for a single physical color-octet component $G_H^a$ ($M_s=2.5\,$TeV, $u=1\,$TeV) for the benchmark choice $y_q=m_q/v$.}
\end{table}

The decay to $t\bar t$ overwhelmingly dominates (BR~$\approx 99.9\%$) because the tree‐level rate \eqref{eq:Gammaqq} scales as $(m_q/v)^2$, so the top channel benefits from the large top Yukawa coupling. Light‐quark modes are suppressed by $(m_{u,d,s}/v)^2\ll10^{-6}$ and are therefore negligible. The loop‐induced $gg$ mode in Eq.~\eqref{eq:Gamma00} is suppressed by the one-loop factor and by the fermion-triangle form factor, and for the benchmark point considered here it remains subleading, with a branching fraction of order $0.01\%$.

\subsubsection{Hadronization and final-state yields for the color octet}

The partial widths and branching fractions of the physical color-octet scalar were used as input to \texttt{Pythia 8} \cite{Sjostrand:2014zea,Bierlich:2022pfr}.  Pythia then performed the hadronization of the primary partonic final states into jets of mesons and baryons, followed by the decays of all unstable resonances into stable photons, leptons, hadrons, and neutrinos.

In these simulations, neutral mesons, especially \(\pi^0\to\gamma\gamma\), provide abundant high-energy photons, while charged mesons and heavy-flavor states decay semileptonically and produce high-energy neutrinos. As a result, photon and neutrino multiplicities from color-octet decays are among the most useful long-range signatures of the TeV-scale scalar sector.

As in the massive-gluon case, the same decay and hadronization pipeline applies independently of how the parent color-octet scalar is produced. The parent scalar may arise from vacuum mismatch, from subcritical thermal freeze-out below \(T_c\), or from non-equilibrium wall--matter transition radiation. These mechanisms differ mainly in the total number of parent particles. For the benchmark parameters considered here, the wall--matter transition-radiation channel gives the dominant parent-particle yield. Therefore, the spectra shown below are normalized to this dominant radiation-production mechanism.

\begin{figure}[htbp]
  \centering
  \includegraphics[width=0.49\textwidth]{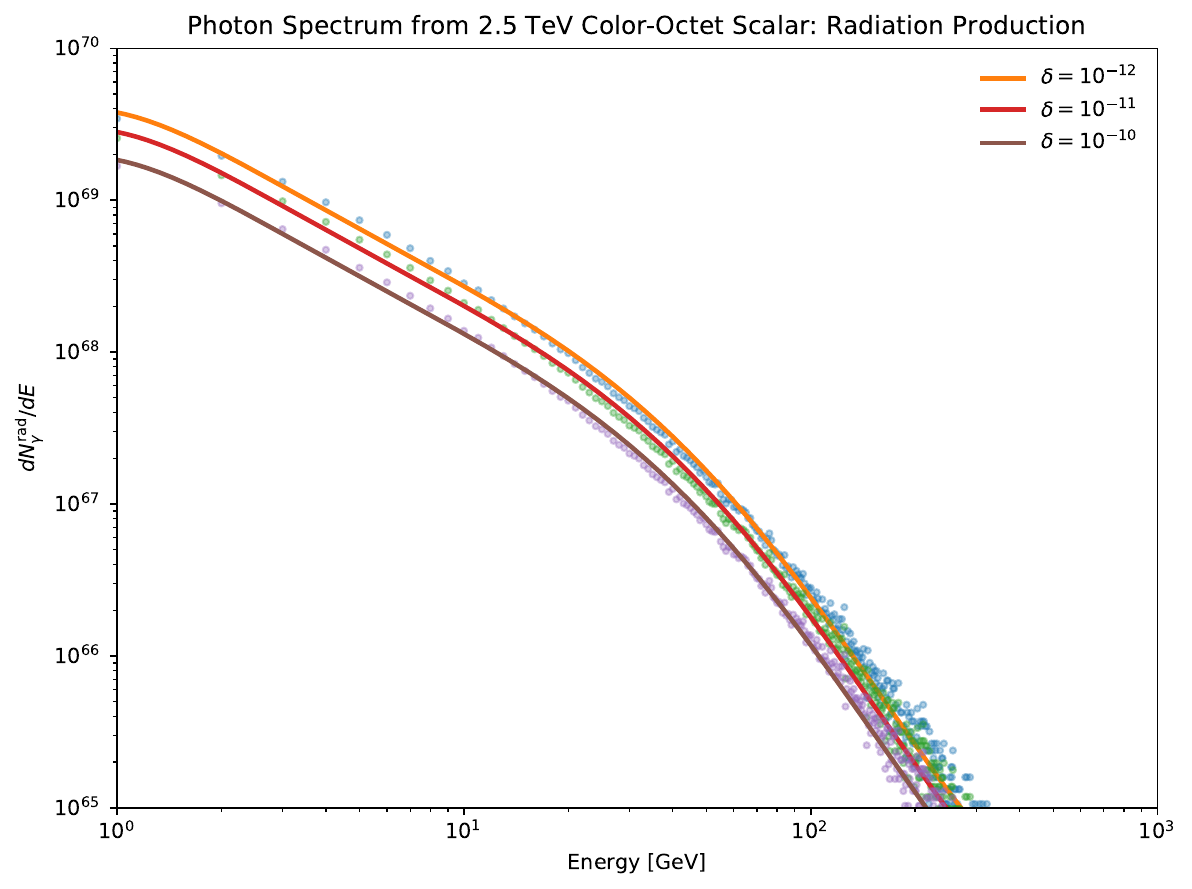}
  \includegraphics[width=0.49\textwidth]{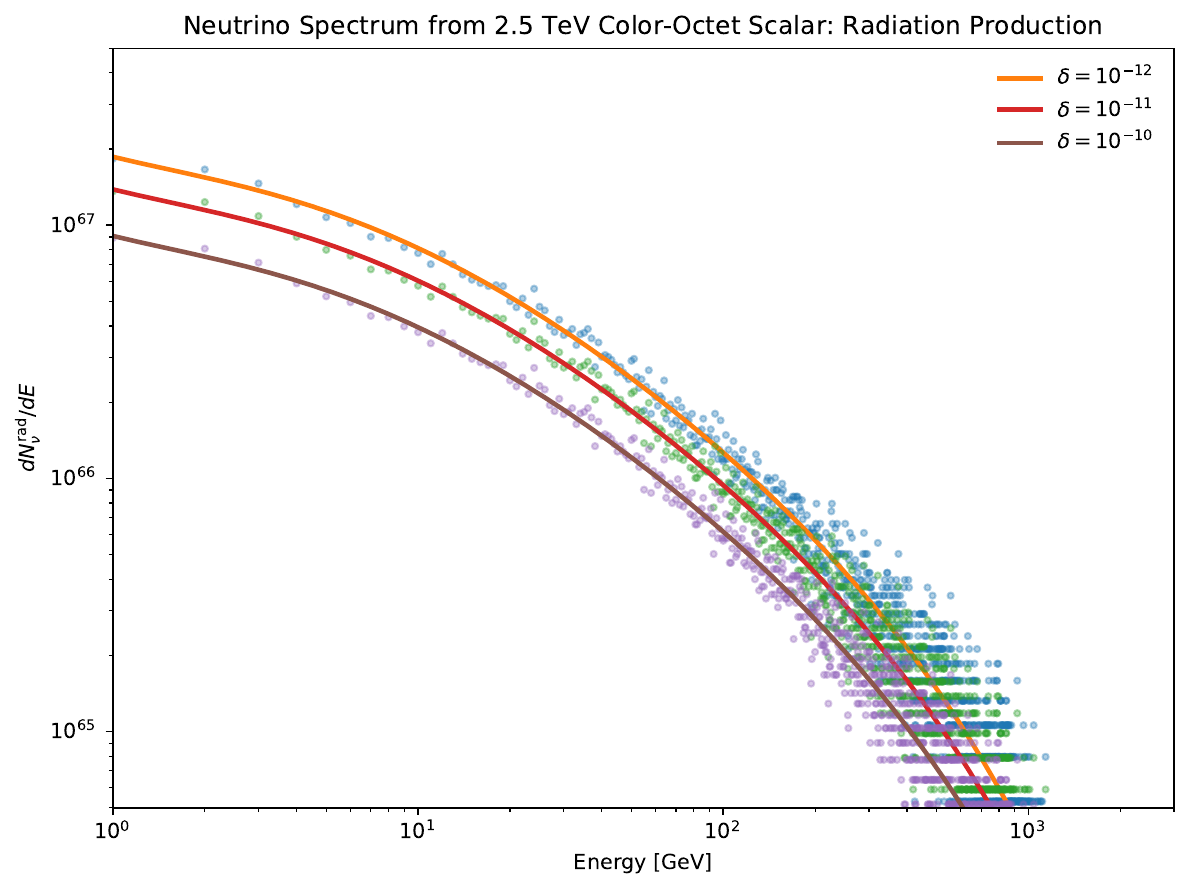}
  \caption{Differential energy spectrum of photons and neutrinos from decays of 2.5\,TeV color-octet scalars produced through the dominant wall--matter transition-radiation channel. The spectral shape is inherited from the per-parent \texttt{Pythia} decay and hadronization spectrum, while the overall normalization is fixed by the transition-radiation yield in Table~\ref{tab:transition-radiation-yields}.}
  \label{fig:photon_spectrum2}
\end{figure}

Figure~\ref{fig:photon_spectrum2} displays the photon and neutrino spectra from the color-octet scalar channel, normalized to the dominant wall--matter transition-radiation production mechanism, which is powered by the enormous amount of ambient matter and energy processed by the relativistic wall as it propagates through the environment. The spectra should therefore be interpreted as rest-frame final-state photon and neutrino yields after parent scalar production, decay, hadronization, and subsequent cascade decays. Since photons and neutrinos propagate over long distances more efficiently than most other stable particles, they provide the most relevant observational signatures of the color-octet scalar sector.
\subsection{Phenomenology of a massive gluon}

In our scenario the $SU(3)_c$ symmetry breaks to $U(2)$, leaving four gluons massless while the remaining four acquire a common mass of order
\begin{equation}
  M_g \simeq \frac{\sqrt3}{2}\,g_s\,\psi_{\rm true}\;\approx\;1~\mathrm{TeV},
\end{equation}
where \(g_s\) is the strong coupling evaluated at the TeV scale, and \(\psi_{\rm true}\simeq1.06~{\rm TeV}\) is the shifted true-vacuum expectation value. In the numerical estimates we round this to \(M_g=1.0~{\rm TeV}\). The remaining four gluons stay massless inside the broken phase and are associated with the unbroken $U(2)$ subgroup rather than with ordinary $SU(3)_c$ QCD.

Inside the true vacuum bubble, the residual unbroken gauge sector mediates the strong dynamics responsible for the formation of bound states of the massive gluons. More precisely, the non-Abelian $SU(2)\subset U(2)$ part is expected to remain asymptotically free and confining in the infrared, while the Abelian $U(1)$ factor mediates long-range Coulombic interactions. At short distances, where asymptotic freedom holds, the inter-gluon potential may be modeled approximately as Coulombic,
\begin{equation}
  V(r)\simeq -\,\frac{C_{\rm eff}\,\alpha_{\rm eff}}{r},
\end{equation}
while at larger separations, confinement induces a linear potential, $\sigma r$, yielding an overall Cornell-type form,
\begin{equation}
  V(r) = -\,\frac{C_{\rm eff}\,\alpha_{\rm eff}}{r} + \sigma r,
\end{equation}
as in the case of heavy quarkonia~\cite{Eichten:1978tg,Brambilla:2010cs}.  
Solving the Schrödinger equation in this hybrid potential gives the typical size of the lowest bound states.  
In the Coulombic region, a Bohr-model estimate analogous to positronium applies because both constituents have the same mass:
\begin{equation}
  r_B \sim \frac{1}{M_g\,\alpha_s(M_g)}.
\end{equation}
For $M_g\simeq1~\mathrm{TeV}$ and $\alpha_s(M_g)\!\simeq\!0.08$,\footnote{For order-of-magnitude estimates we use the Standard-Model QCD coupling evaluated at the TeV scale as a proxy for the broken-phase non-Abelian coupling; a dedicated renormalization-group treatment of the residual U(2) theory is left for future work.} one obtains
$r_B\!\approx\!2.5\times10^{-18}\,\mathrm{m}$,
which is about an order of magnitude larger than the Compton wavelength of the gluon
$\lambda_C=\hbar c/M_g\!\approx\!2\times10^{-19}\,\mathrm{m}$.
At this distance, the Yukawa suppression factor for exchange of the heavy broken gluons is
\begin{equation}
  e^{-M_g r_B} = e^{-1/\alpha_s(M_g)} \approx e^{-12.5} \approx 4\times10^{-6},
\end{equation}
demonstrating that direct exchange of the heavy broken gluons between such bound states is exponentially small.  
Thus, the heavy-gluon dynamics are expected to be dominated by the unbroken gauge interaction, whose potential determines the structure and size of the bound states.  
This is also consistent with the effective-field-theory expectation that sufficiently heavy broken-sector gauge bosons decouple from long-distance dynamics, in the spirit of Appelquist--Carazzone~\cite{Appelquist:1974tg}, and with the expectation that the confinement radius is set by the infrared scale of the unbroken non-Abelian sector as argued by Goncharov~\cite{Goncharov:2012yx}.  

This behavior is analogous to the hierarchy observed in heavy-quark systems such as charmonium and bottomonium, where the bound-state size scales as $1/(m_Q\alpha_s)$ and the potential interpolates smoothly between Coulombic and confining regimes. In the present scenario, the corresponding heavy-gluon bound states are considerably more compact, but their sizes remain controlled by the same parametric scaling, ensuring that their formation and decay dynamics mirror those of conventional heavy quarkonia at proportionally higher mass scales.

In contrast, outside the bubble—the false vacuum corresponding to the Standard Model QCD phase—the confinement scale remains
\(\Lambda_{\rm QCD}^{\rm SM}\!\sim\!100~\mathrm{MeV}\).
There, the typical hadron radius made up of light quarks is \(r_{\rm hadron}^{\rm SM}\!\sim\!1~\mathrm{fm}\), while the TeV-scale gluon has a range \(\lambda_C\!\sim\!10^{-18}\,\mathrm{m}\!\ll\!r_{\rm hadron}^{\rm SM}\).
Consequently, the Yukawa factor \(e^{-M_g r}\) becomes exponentially small, and the heavy gluon completely decouples from low-energy hadronic physics, as expected. Outside the bubble, ordinary $SU(3)_c$ QCD is recovered, whereas inside the true vacuum the long-distance dynamics are governed by the residual $U(2)$ gauge theory and therefore differ qualitatively from standard QCD.

It is useful to note why we do not include decays of a single massive gluon
into the massless gluons of the unbroken \(U(2)\) sector. Let \(i,j\) denote
unbroken generators and \(\hat a\) denote broken generators. Since the
unbroken generators close among themselves, \([H,H]\subset H\) with
\(H=U(2)\), the structure constants with one broken and two unbroken indices
vanish,
\begin{equation}
f^{\hat aij}=0.
\end{equation}
Thus the triple-gauge vertex containing one massive broken gluon and two massless unbroken gluons is absent at tree level. Vertices involving one massless and two massive gluons can exist, but a decay \(G_{\hat a}\to G_{\hat b}+g_i\) is kinematically forbidden for degenerate massive gluons, apart from the zero-measure soft limit. Therefore the leading
phenomenologically relevant decay channel of a single massive gluon is into quark--antiquark pairs through the usual QCD interaction
\begin{equation}
  \mathcal{L}_{\rm int} = g_s\,\bar{q}_i\gamma^\mu T^a_{ij} q_j\,G^a_\mu.
\end{equation}
In this framework, the tree-level partial width for $G\to q\bar q$ can be derived from color, spin, and two-body phase space (see Appendix~\ref{App-D}). Summing over final-state colors and spins and averaging over the three initial polarizations of a massive vector yields
\begin{equation}\label{eq:partial_width}
  \Gamma(G\to q\bar q) 
  = \frac{\alpha_s}{6}\,M_g\,
    \Bigl(1+\frac{2m_q^2}{M_g^2}\Bigr)
    \sqrt{1-\frac{4m_q^2}{M_g^2}}\,.
\end{equation}
The coefficient \(1/6\) arises from the fundamental color factor \(T(R)=\tfrac12\), the average over three vector polarizations \((2J{+}1=3)\), and the standard \(1\!\to\!2\) phase-space normalization.

The kinematic factor $\sqrt{1-4m_q^2/M_g^2}$ ensures that for $M_g<2m_q$ the decay is forbidden, rising smoothly from zero at the threshold $M_g=2m_q$. The prefactor $(1+2m_q^2/M_g^2)$ gives a modest enhancement near threshold (notably for top). Expanding near threshold,
\begin{equation}
  \Gamma\propto\alpha_s M_g\sqrt{\delta},
  \quad\delta=1-\frac{4m_q^2}{M_g^2}.
\end{equation}

In the limit where the gluon mass greatly exceeds all quark masses ($M_g\gg m_q$), mass-dependent corrections become negligible and the partial widths simplify to
\begin{equation}
  \Gamma(G\to q\bar q)\simeq\frac{\alpha_s}{6}\,M_g.
\end{equation}
Since every quark flavor above threshold contributes equally, the branching fractions approach a universal value:
\begin{equation}
  \mathrm{BR}(G\to q\bar q)\;\simeq\;\frac{1}{N_f},
\end{equation}
where $N_f$ is the number of kinematically open flavors. For $M_g>2m_t$, one has $N_f=6$, yielding $\mathrm{BR}\approx16.7\%$ for each quark species.

\subsubsection{Branching ratio and decay width at \texorpdfstring{$M_g \sim 1$\,TeV}{Mg ~ 1 TeV}.}

The branching ratios and decay widths of the massive gluons to different quarks are listed in Table~\ref{Tab:GluonDecay}, using $\alpha_s(1~\text{TeV})=0.09$ and $m_t=173$ GeV.
\begin{table}[htbp]
  \centering
  \begin{tabular}{@{}lcc@{}}
    \toprule
    Gluon mass $M_g = 1$~TeV & & \\
    Total decay width: $89.91$~GeV & & \\
    \midrule
    Quark      & Decay Width (GeV) & BR (\%) \\
    \midrule
    up         & $15.00$           & $16.68$ \\
    down       & $15.00$           & $16.68$ \\
    strange    & $15.00$           & $16.68$ \\
    charm      & $15.00$           & $16.68$ \\
    bottom     & $15.00$           & $16.68$ \\
    top        & $14.91$           & $16.59$ \\
    \bottomrule
  \end{tabular}
  \caption{\label{Tab:GluonDecay}Partial widths and branching ratios for a $1$ TeV massive gluon at tree level, using Eq.~\eqref{eq:partial_width} with $\alpha_s=0.09$.}
\end{table}

Summing over the six kinematically accessible quarks yields
\[
  \Gamma_{\rm tot}\simeq 89.91\;\mathrm{GeV},
  \qquad \frac{\Gamma}{M_g}\sim 0.09,
\]
for a gluon mass of 1~TeV. The branching ratio of the top channel rises from its threshold-suppressed value and approaches the democratic limit of $16.7\%$ as the gluon mass increases and the kinematic suppression becomes negligible.

\subsubsection{Hadronization and final-state yields}

The six partial widths and branching ratios calculated for a $1$ TeV massive gluon---with a total width of $\simeq 89.9$ GeV and each light-flavor channel at roughly $16.7\%$ (top quark at $16.6\%$)---were used as the hard-process input in \texttt{Pythia\,8} \cite{Sjostrand:2014zea, Bierlich:2022pfr}.  Pythia then handled the fragmentation of every \(q\bar q\) pair into hadrons, including mesons and baryons, and simulated the decays of all unstable resonances.

Inside each jet, neutral pions, especially \(\pi^0\to\gamma\gamma\), provide abundant high-energy photons, while charged pions, kaons, and heavy-flavor hadrons undergo semileptonic decays that yield high-energy neutrinos. As a result, the photon and neutrino counts per massive-gluon decay exceed those of most other stable particle species. This makes high-energy photons and neutrinos the most useful long-range messengers of TeV-scale massive-gluon production.

As discussed above, the same decay and hadronization pipeline applies to massive gluons produced by vacuum mismatch, by subcritical thermal freeze-out, or by wall--matter transition radiation. These mechanisms differ mainly in the normalization of the parent massive-gluon population. For the benchmark parameters considered here, the wall--matter transition-radiation channel gives the largest parent-particle yield. Therefore, in the spectra shown below we focus on the photon and neutrino signal normalized to this dominant radiation-production mechanism.

\begin{figure}[htb]
  \centering
  \includegraphics[width=0.50\textwidth]{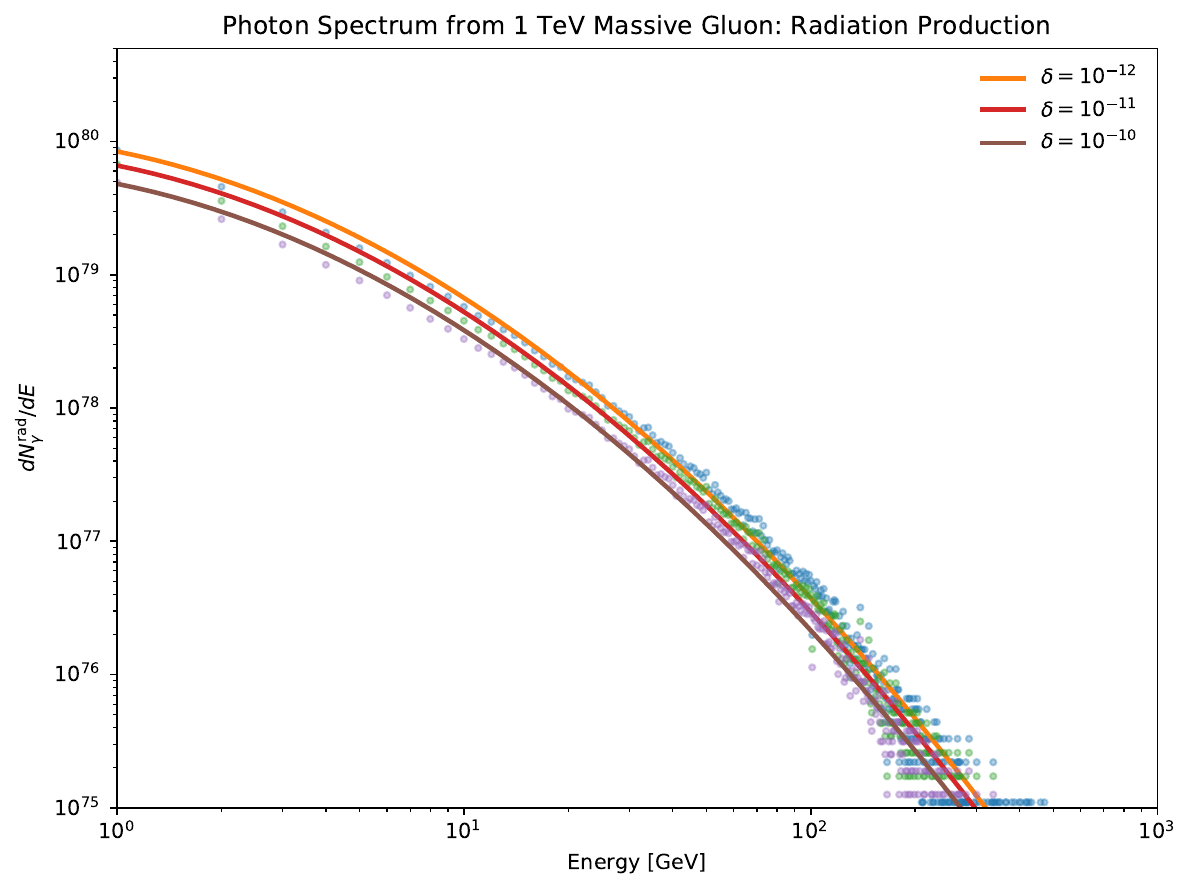}
  \includegraphics[width=0.49\textwidth]{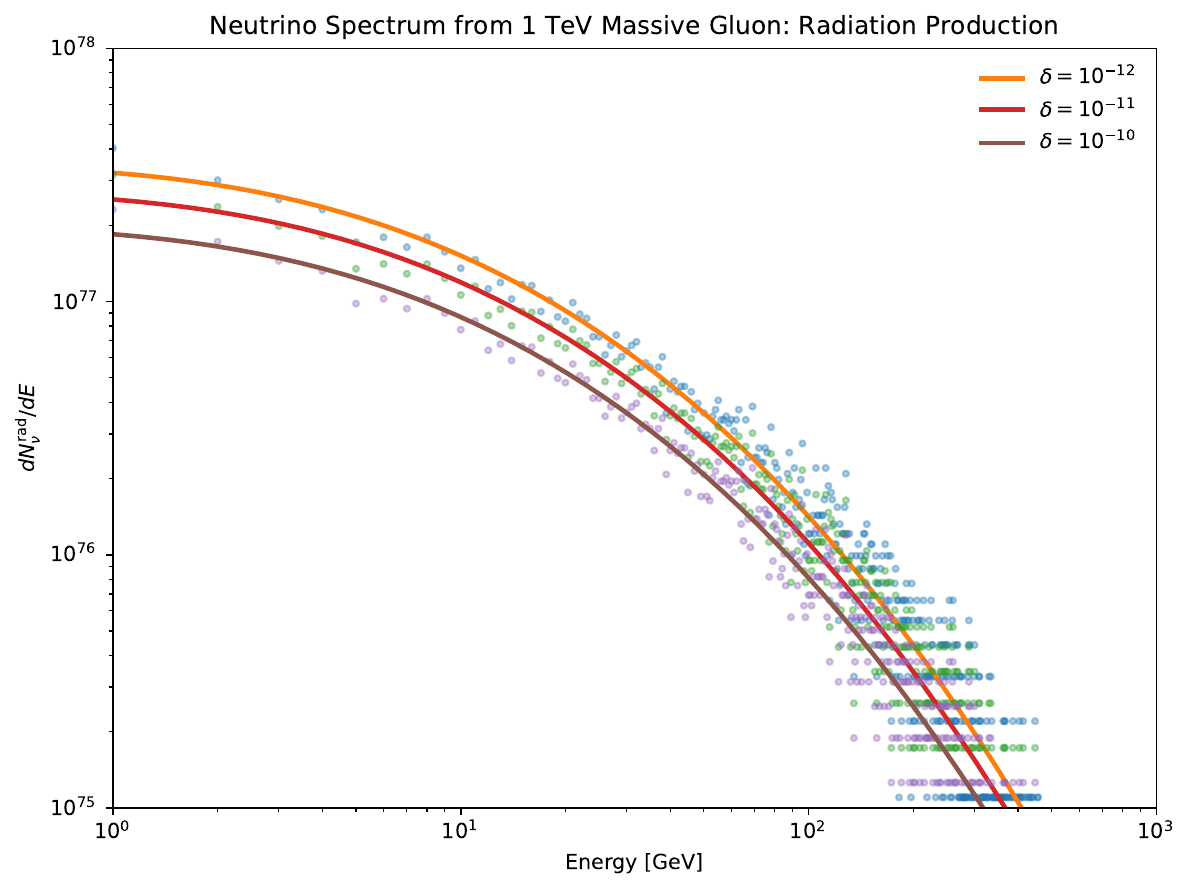}
  \caption{Differential energy spectra of photons and neutrinos from decays of $1$ TeV massive gluons produced through the dominant wall--matter transition-radiation channel. The spectral shape is inherited from the per-parent \texttt{Pythia} decay and hadronization spectrum, while the overall normalization is fixed by the transition-radiation yield in Table~\ref{tab:transition-radiation-yields}.}
  \label{fig:photon_spectrum3}
\end{figure}

Figure~\ref{fig:photon_spectrum3} displays the resulting photon and neutrino spectra for the $1$ TeV massive gluon, normalized to the dominant wall--matter transition-radiation production channel. The vacuum-mismatch and subcritical thermal channels would lead to the same per-parent decay spectra, but with smaller overall normalizations for the benchmark parameters used here.

The spectra in Figs.~\ref{fig:photon_spectrum2}, and \ref{fig:photon_spectrum3} describe the photon and neutrino yields at the site of production. To obtain the spectra observed at Earth, propagation effects must be included. The corresponding particle flux, i.e. the number of particles per unit area per unit time, is diluted by a factor of \(1/[4\pi d^2(1+z)]\), where \(d\) is the physical distance between the source and the observer and \(z\) is the source redshift. The factor \(1+z\) accounts for cosmological time dilation: two particles emitted a time interval \(\Delta t\) apart are measured an interval \((1+z)\Delta t\) apart. The individual particle energies are also redshifted by the same factor. For high-energy photons, propagation over cosmological distances may also include attenuation through pair production on the extragalactic background light and the CMB, followed by electromagnetic cascading. The simple geometric factor should therefore be understood as the geometric and redshift dilution only. Neutrinos are less affected by such propagation effects.

\section{Photon or neutrino signal lead time}
\label{sec:signal_leadtime-su3c}

If bubble walls in the $SU(3)_c$ scenario propagate at slightly subluminal speeds, any secondary radiation (photons or neutrinos from colored scalar or massive gluons) will arrive earlier than the wall itself. Such particles may thus act as a cosmological ``advance warning'' of the approaching bubble. We now quantify this delay for a source situated at a distance of $1$ billion light years.

\paragraph{Cosmological inputs and arrival delay}

For the benchmark estimate considered here, a full cosmological treatment is not necessary.  
We take a source at a distance
\begin{equation}
D = 10^9\ {\rm ly},
\end{equation}
which corresponds to a modest redshift, $z\simeq 0.07$, in a flat $\Lambda$CDM cosmology. At such low redshift, the difference between the exact cosmological result and the flat-space estimate is negligible for our purposes, so we use the Minkowski approximation throughout this subsection. If the bubble wall propagates at speed
\begin{equation}
v=(1-\delta)c,
\qquad \delta\ll 1,
\end{equation}
then the arrival delay relative to photons or neutrinos is
\begin{equation}
\Delta t \simeq D\!\left(\frac{1}{v}-\frac{1}{c}\right)
= \frac{\delta}{1-\delta}\,\frac{D}{c}
\simeq \delta\,\frac{D}{c}.
\label{eq:delay-flat-su3c}
\end{equation}
For $D=10^9$ light years, this gives the lead times listed in Table~\ref{tab:photon_delay_su3c}.

\begin{table}[h!]
\centering
\begin{tabular}{|ccc|}
\hline
\textbf{Velocity Deficit $\delta$} & \textbf{Distance (ly)} & \textbf{Time Delay} \\
\hline
$1.0\times10^{-10}$ & $1.0\times10^9$ & 36 d, 12 h, 36 m, 0 s \\
$1.0\times10^{-11}$ & $1.0\times10^9$ & 3 d, 15 h, 39 m, 36 s \\
$1.0\times10^{-12}$ & $1.0\times10^9$ & 0 d, 8 h, 45 m, 57.6 s \\
\hline
\end{tabular}
\caption{Photon/neutrino lead times for bubble walls with subluminal deficits $\delta$ created at a distance of $10^9$ light years from us, using the flat-space approximation $\Delta t \simeq \delta D/c$. Even for $\delta=10^{-12}$, the lead time is several hours, while larger deficits extend the warning to days or weeks.}
\label{tab:photon_delay_su3c}
\end{table}

Thus, in the SU(3)\(_c\) model, even minuscule velocity deficits can produce noticeable precursor signals. Depending on 
$\delta,$ photons and neutrinos produced in the decay and hadronization cascades of the colored scalar or massive gluons could reach observers hours to weeks before the bubble wall. We emphasize, however, that these benchmark lead times are not unique. They scale with both the velocity deficit and the distance to the nucleation event, $\Delta t\simeq \delta D/c,$ and can therefore be longer for different choices of the color-breaking scale, friction coefficient, terminal velocity, or for bubbles nucleated at larger distances.

\section{Conclusions} 

In this work, we investigated the potential cosmological signatures stemming from a catastrophic phase transition in which the strong nuclear force, described by the $SU(3)_c$ gauge symmetry, becomes spontaneously broken. While the $SU(3)_c$ symmetry defines the fundamental structure of our present-day universe and is essential for life as we know it, there is no fundamental principle guaranteeing its eternal persistence. To explore the observational consequences of such an event, we constructed a phenomenological model incorporating a new colored scalar field responsible for the symmetry breaking. The potential of this field facilitates a first-order phase transition, driven by the nucleation and expansion of true-vacuum bubbles inside the surrounding false vacuum of our current $SU(3)_c$-symmetric state.

We analyzed three distinct sources of particles associated with the expanding wall. The first is direct vacuum-mismatch production. This is the most controlled nonthermal channel: as the wall passes, the scalar and gauge-field modes are matched between the unbroken and broken phases, producing massive scalar excitations and massive gluons. This contribution provides a conservative baseline signal, but it is strongly limited by the exponential dependence on the wall acceleration and becomes inefficient once terminal motion is reached.

The second source is thermal production from frictional dissipation. Interactions between the wall and the ambient medium transfer energy into a shocked layer. We showed, however, that this channel must be treated together with the finite-temperature phase diagram of the $SU(3)_c$-breaking scalar. For the benchmark studied here, the critical temperature is
\begin{equation}
T_c\simeq0.256~{\rm TeV},
\end{equation}
which is below both the massive-gluon mass and the physical color-octet scalar mass. Therefore an equilibrated thermal bath hot enough to make these broken-phase massive particles ultrarelativistic would restore $SU(3)_c$, and cannot consistently be described as a gas of massive broken-phase gluons and scalar octets. We therefore formulated a subcritical thermal freeze-out estimate below $T_c$, where the heavy states can still be produced but only with Boltzmann suppression. The resulting yields are parametrized by an efficiency factor $f_{\rm th}$, which encodes the fraction of dissipated wall energy that passes through a subcritical broken-phase thermal stage.

The third source is non-equilibrium wall--matter transition radiation. Ambient matter crossing the relativistic $SU(3)_c\to U(2)$ wall is highly boosted in the wall frame. Since the wall changes the gauge spectrum and the color-field configuration of the incident matter, it can induce radiation of massive broken-phase gluons and scalar excitations. This process is not governed by a plasma temperature and therefore avoids the finite-temperature restoration objection that applies to a hot equilibrated thermal bath. It can also continue after the wall reaches terminal velocity, since it only requires ambient matter to cross the wall. Its normalization is therefore controlled by the amount of matter swept up by the bubble and by the microscopic transition-radiation efficiencies.

After production, the unstable massive gluons and color-octet scalar decay into quarks and gluons, whose subsequent hadronization and decay chains were modeled with \texttt{Pythia}. The final long-range messengers are high-energy photons and neutrinos. The spectra derived in this work should therefore be interpreted as parent-particle decay and hadronization spectra, with different production mechanisms providing different normalizations for the parent population. Vacuum mismatch gives a conservative nonthermal baseline; subcritical thermal freeze-out can enhance the signal if the efficiency $f_{\rm th}$ is not too small; and wall--matter transition radiation can be important in sufficiently dense environments or over sufficiently long propagation distances.

If the bubble wall travels exactly at the speed of light, no signal can overtake it. However, interactions with matter, radiation, or self-produced particles can slow the wall to a subluminal terminal velocity. In that case, photons and neutrinos produced near the wall may arrive before the wall itself. Even extremely small velocity deficits can generate macroscopic lead times over cosmological distances. For a source at $D=10^9$ light years, the benchmark deficits $\delta=10^{-12}$--$10^{-10}$ correspond to warning times ranging from several hours to several weeks.

We emphasize that the late-time signal is not universal. The vacuum-mismatch channel is controlled by the microscopic wall dynamics and provides a minimal baseline. The subcritical thermal channel depends on how efficiently dissipated energy freezes out below the restoration temperature. The wall--matter transition-radiation channel depends on the ambient matter density, propagation distance, and microscopic radiation efficiencies. Thus the brightness of a bubble is environment-dependent: it may be dim in cosmic voids but much brighter when sweeping through denser matter. If a correlated high-energy photon and neutrino signal with the spectral and temporal features described here were ever observed, it could be interpreted as a possible precursor of a late-time color-breaking phase transition.    

\begin{acknowledgments}
The authors are grateful to Jure Zupan, D.C. Dai and Manuel Szewc for carefully reviewing the manuscript and for their valuable comments and suggestions. We also thank Michael Peskin for discussions which motivated us to consider
the wall--matter transition-radiation channel. AS and DS are partially supported by the U.S. National
Science Foundation, under the Grant No. PHY-2310363. AS is also supported by the Grant No. NSF OAC-2417682. L.C.R. W is supported by the U.S. Department of Energy grant DE-SC0011784.
\end{acknowledgments}

\appendix
\section{Appendix~A: Symmetry breaking patterns from the adjoint potential}\label{App-A}

In this appendix we expand upon the discussion of the scalar potential presented in the Model section [see~\ref{Model}] of the manuscript.  
Our goal is to analyze in detail how the vacuum alignment of an $SU(3)_c$ adjoint Higgs determines the residual gauge symmetry, and to count the number of massless and massive gluons in each case.  
We emphasize the role of the cubic invariant, which lifts the angular degeneracy present in the purely quadratic-plus-quartic potential and selects the $U(2)$-type vacuum alignment relevant for the main text.

\smallskip
\noindent
To avoid confusion with the effective one-field parameters used in the canonically normalized tunneling coordinate in Sec.~\ref{Model}, the coefficients in this appendix are denoted by
\(\mu_\Phi^2,\lambda_{1\Phi},\lambda_{2\Phi}\). These are the trace-level coefficients of the adjoint potential before projection onto the one-dimensional order-parameter direction.

\subsection*{Lagrangian and potential}

The gauge--scalar sector is given by
\begin{equation}
\mathcal{L}
= -\frac{1}{2}\text{Tr} F_{\mu\nu}F^{\mu\nu}
+ \frac{1}{2}\text{Tr} (D_\mu\Phi)(D^\mu\Phi)
- V(\Phi),
\end{equation}
where
\begin{equation}
F_{\mu\nu}=\partial_\mu A_\nu-\partial_\nu A_\mu+i g [A_\mu,A_\nu],
\qquad
D_\mu\Phi=\partial_\mu\Phi+i g [A_\mu,\Phi].
\end{equation}
The scalar potential is
\begin{equation}
V(\Phi)
=
\frac{\mu_\Phi^2}{4}\,\text{Tr}(\Phi^2)
+
\frac{\lambda_{1\Phi}}{16}\,\bigl(\text{Tr}\Phi^2\bigr)^2
+
\frac{\lambda_{2\Phi}}{6}\,\text{Tr}(\Phi^3)
+ V_0,
\qquad
\lambda_{1\Phi}>0,
\end{equation}
with the constant $V_0$ set to zero henceforth.

\subsection*{Diagonal VEV and two--variable reduction}

We parameterize the vacuum as
\begin{equation}
\langle\Phi\rangle=v=\mathrm{diag}(a,b,c),
\qquad
a+b+c=0.
\end{equation}
The invariants become
\begin{align}
\text{Tr}(v^2)&=a^2+b^2+c^2=:S_2, \\
\text{Tr}(v^3)&=a^3+b^3+c^3=3abc,
\end{align}
so the potential on the VEV subspace reads
\begin{equation}
V(v)
=
\frac{\mu_\Phi^2}{4}(a^2+b^2+c^2)
+
\frac{\lambda_{1\Phi}}{16}(a^2+b^2+c^2)^2
+
\frac{\lambda_{2\Phi}}{2}\,abc.
\end{equation}
Eliminating $c=-(a+b)$ and defining
\begin{equation}
Q:=a^2+b^2+ab
\qquad
(\Rightarrow S_2=2Q),
\qquad
abc=-ab(a+b),
\end{equation}
we obtain the convenient two-variable form
\begin{equation}
V(a,b)
=
\frac{\mu_\Phi^2}{2}\,Q
+
\frac{\lambda_{1\Phi}}{4}\,Q^2
-
\frac{\lambda_{2\Phi}}{2}\,ab(a+b).
\label{A.9}
\end{equation}

\subsection*{Case I: cubic present ($\lambda_{2\Phi}\neq 0$) $\Rightarrow U(2)$}

From \eqref{A.9},
\begin{align}
\frac{\partial V}{\partial a}
&=
\left(\frac{\mu_\Phi^2}{2}+\frac{\lambda_{1\Phi}}{2}Q\right)(2a{+}b)
-\frac{\lambda_{2\Phi}}{2}\,b(2a{+}b)
=0,
\tag{A.10}\\
\frac{\partial V}{\partial b}
&=
\left(\frac{\mu_\Phi^2}{2}+\frac{\lambda_{1\Phi}}{2}Q\right)(2b{+}a)
-\frac{\lambda_{2\Phi}}{2}\,a(2b{+}a)
=0.
\tag{A.11}
\end{align}
Equivalently, after multiplying by $2$, these may be written as
\begin{align}
(2a+b)\bigl(\mu_\Phi^2+\lambda_{1\Phi}Q-\lambda_{2\Phi} b\bigr)&=0,
\tag{A.10'}\\
(2b+a)\bigl(\mu_\Phi^2+\lambda_{1\Phi}Q-\lambda_{2\Phi} a\bigr)&=0.
\tag{A.11'}
\end{align}
The cubic invariant lifts the angular degeneracy of the purely quadratic-plus-quartic potential. Up to permutations of the three eigenvalues, the relevant stationary branch has
\begin{equation}
a=b,
\qquad
c=-2a,
\end{equation}
for which two eigenvalues coincide. On this branch \(Q=3a^2\), and the nontrivial stationary condition becomes
\begin{equation}
\mu_\Phi^2+3\lambda_{1\Phi}a^2-\lambda_{2\Phi}a=0.
\end{equation}
This is precisely the vacuum alignment relevant to the benchmark used in the main text, and it leads to the breaking pattern
\begin{equation}
SU(3)\ \longrightarrow\ U(2)\simeq [SU(2)\times U(1)]/\mathbb{Z}_2,
\end{equation}
with
\begin{equation}
4~\text{massless gluons},
\qquad
4~\text{massive gluons}.
\label{A.14}
\end{equation}

\subsection*{Case II: cubic absent ($\lambda_{2\Phi}=0$) $\Rightarrow$ $U(1)\times U(1)$ for generic vacua with $\mu_\Phi^2<0$}

With $\lambda_{2\Phi}=0$ the adjoint potential reads
\begin{equation}
V(a,b)
=
\frac{\mu_\Phi^2}{2}\,Q
+
\frac{\lambda_{1\Phi}}{4}\,Q^2,
\qquad
Q \equiv a^2+b^2+ab,
\label{eq:case2V}
\end{equation}
where we take $\lambda_{1\Phi}>0$ so that the potential is bounded from below. The stationarity conditions are
\begin{align}
\frac{\partial V}{\partial a}
&=
\frac12(2a{+}b)\,\bigl[\mu_\Phi^2+\lambda_{1\Phi} Q\bigr]
=0,
\label{eq:dVda}\\
\frac{\partial V}{\partial b}
&=
\frac12(2b{+}a)\,\bigl[\mu_\Phi^2+\lambda_{1\Phi} Q\bigr]
=0.
\label{eq:dVdb}
\end{align}

\paragraph{On the sign of $\mu_\Phi^2$.}

If $\mu_\Phi^2>0$, then
\[
V=\frac{\mu_\Phi^2}{2}Q+\frac{\lambda_{1\Phi}}{4}Q^2\ge 0
\]
with $V=0$ only at $Q=0\Rightarrow a=b=0$. In this case the only local minimum is the trivial vacuum \(a=b=c=0\), and the full \(SU(3)\) remains unbroken. Thus nontrivial vacua arise only if \(\mu_\Phi^2<0\).

\paragraph{Nontrivial vacua for $\mu_\Phi^2<0$.}

For $\mu_\Phi^2<0$, the minimum is determined by the radial condition
\begin{equation}
\mu_\Phi^2+\lambda_{1\Phi}Q=0
\qquad\Longrightarrow\qquad
Q=-\frac{\mu_\Phi^2}{\lambda_{1\Phi}}.
\end{equation}
This fixes only the radius in the \((a,b)\)-plane, while the angular direction remains undetermined. This angular degeneracy is the accidental flat direction present when the cubic invariant is absent.

It is often convenient to introduce the angular ratio
\begin{equation}
r\equiv \frac{b}{a}.
\end{equation}
For generic
\begin{equation}
r\neq 1,\,-2,\,-\frac12,
\end{equation}
the three eigenvalues
\[
a,\qquad b,\qquad c=-(a+b)
\]
are all distinct. The special case
\begin{equation}
a+b=0
\qquad\Rightarrow\qquad
b=-a,\qquad c=0
\end{equation}
is simply one representative point on the same vacuum manifold; in this case \(Q=a^2\), and the radial condition becomes
\begin{equation}
\mu_\Phi^2+\lambda_{1\Phi}a^2=0
\quad\Longrightarrow\quad
a^2=-\frac{\mu_\Phi^2}{\lambda_{1\Phi}},
\qquad
(\mu_\Phi^2<0).
\label{eq:branchI}
\end{equation}

\subsection*{Residual symmetry and gauge--boson masses}

Let the three adjoint eigenvalues be \((a,b,c)\), with \(a+b+c=0\). The tree-level gauge masses follow from the adjoint covariant derivative. Up to the normalization convention chosen for the adjoint vacuum expectation value, the off-diagonal gauge-boson masses scale as
\begin{equation}
m^2(E_{ij}) \propto g^2\,(a_i-a_j)^2
\quad
(i\neq j),
\qquad
m^2(\text{Cartan})=0,
\label{eq:gaugemasses}
\end{equation}
where the six off-diagonal generators \(E_{ij}\) acquire masses proportional to the squared differences of the adjoint eigenvalues, while the two diagonal Cartan generators remain massless.

For generic, nondegenerate eigenvalues one obtains the breaking pattern
\[
SU(3)\ \longrightarrow\ U(1)\times U(1),
\qquad
6~\text{massive gluons},
\quad
2~\text{massless Cartan gluons}.
\]
This situation corresponds to the generic angular branch
\[
r\neq 1,\,-2,\,-\frac12,
\]
and also includes the representative point \(b=-a\), with \(c=0\), since all three eigenvalues are then pairwise distinct.

In the special degenerate cases
\[
r=1,\qquad r=-2,\qquad r=-\frac12,
\]
two eigenvalues coincide and the symmetry is enhanced to
\[
SU(3)\ \longrightarrow\ U(2),
\]
yielding four massive and four massless gluons.

\begin{table}
\begin{center}
\scriptsize
\renewcommand{\arraystretch}{1.2}
\begin{tabular}{@{}lcccc@{}}
\toprule
Potential choice & Vacuum alignment & Unbroken $H$ & Massive/Massless gluons & Scalar spectrum \\
\midrule
$\lambda_{2\Phi}\neq 0$ (cubic present) 
& $a=b\neq c$ 
& $U(2)$ 
& $4$ massive, $4$ massless 
& $4$ physical massive, $4$ eaten \\
$\lambda_{2\Phi}=0$, $\mu_\Phi^2>0$ 
& $a=b=c=0$ (trivial) 
& $SU(3)$ 
& $0$ massive, $8$ massless 
& $8$ massive, $0$ eaten \\
$\lambda_{2\Phi}=0$, $\mu_\Phi^2<0$ 
& $a\neq b\neq c$ (generic) 
& $U(1)\times U(1)$ 
& $6$ massive, $2$ massless 
& $1$ radial massive, $1$ flat, $6$ eaten \\
\bottomrule
\end{tabular}
\end{center}
\caption{
Summary of vacuum alignments and residual gauge symmetries arising from the
adjoint scalar potential of $SU(3)_c$, together with the associated gluon and
scalar spectra. The scalar counts refer to the adjoint scalar degrees of
freedom after symmetry breaking, with Goldstone modes counted as eaten by the
corresponding massive gauge bosons. In the $\lambda_{2\Phi}=0$, $\mu_\Phi^2<0$
case, the extra uneaten scalar is a tree-level flat direction associated with
the accidental angular degeneracy of the potential.
}
\label{tab:appendixA-breaking-patterns}
\end{table}

\paragraph{Order of the transition.}

The breaking pattern and the order of the transition are logically distinct. 
For \(\lambda_{2\Phi}\neq 0\), the cubic invariant \(\mathrm{Tr}(\Phi^3)\) generically creates a barrier between the symmetric and broken phases along the order-parameter direction, yielding a first-order transition for the parameter range used in the main text. The thin-wall bounce treatment then applies when the two vacua are nearly degenerate.

By contrast, for \(\lambda_{2\Phi}=0\) the potential reduces to
\[
V=\frac{\mu_\Phi^2}{2}Q+\frac{\lambda_{1\Phi}}{4}Q^2,
\qquad
Q=a^2+b^2+ab,
\]
with \(\lambda_{1\Phi}>0\). There is no tree-level barrier: as \(\mu_\Phi^2\) passes through zero, the minimum moves continuously from
\[
Q=0
\]
to
\[
Q=-\frac{\mu_\Phi^2}{\lambda_{1\Phi}}.
\]
Thus, at tree level the \(\lambda_{2\Phi}=0\) branch is second order, or continuous. A generic adjoint vacuum expectation value breaks \(SU(3)\to U(1)\times U(1)\), while along the degenerate alignments \(a=b\), \(a=c\), or \(b=c\), the unbroken subgroup is enhanced to \(U(2)\).

When the adjoint scalar acquires a VEV aligned with the eighth Gell--Mann generator, the color group breaks as
\[
SU(3)_c\to U(2)\simeq SU(2)_{rg}\times U(1)_X,
\]
where the unbroken generators \(\{T_{1,2,3},T_8\}\) act block-diagonally on the \((r,g)\) and \(b\) color directions. As a result, the red--green sector forms an \(SU(2)_{rg}\) doublet while the blue quark is an \(SU(2)\) singlet carrying only \(U(1)_X\) charge. The four off-diagonal generators \(\{T_{4-7}\}\) do not commute with \(T_8\), so their gauge bosons acquire masses of order the color-breaking scale and decouple from infrared physics.

Below the breaking scale the couplings of the unbroken subgroup run differently: \(SU(2)_{rg}\) remains asymptotically free and becomes confining at low energies, while the Abelian \(U(1)_X\) coupling weakens like QED. Consequently, the dynamics inside the broken-color phase resemble a confining two-color gauge theory coupled to a weak long-range \(U(1)\): \((r,g)\) quarks form \(SU(2)\) mesons and diquark baryons, whereas blue quarks are singlets under the confining \(SU(2)_{rg}\) factor, so the usual three-color baryon picture is no longer operative in the broken phase.

The massive gluons do not bind via massive-gluon exchange, which is exponentially suppressed at distances larger than the inverse heavy-gluon mass. However, they can form bound states because they remain charged under the unbroken \(SU(2)_{rg}\times U(1)_X\) subgroup and therefore continue to participate in the residual confining \(SU(2)\) dynamics. In summary, the broken phase realizes a rich \(U(2)\) gauge structure with qualitatively new spectroscopy, modified running, and non-standard hadronization patterns; a detailed quantitative analysis of these features, including spectrum, potentials, and RG flows, will be developed in a follow-up paper.

\section{Appendix~B: Quartic trace identity for \texorpdfstring{$SU(3)$}{SU(3)} adjoint fields}
\label{App-B}

We establish the relation
\begin{equation}
(\Tr M^{2})^{2}=2\,\Tr M^{4},
\label{eq:main-identity}
\end{equation}
for $M=B_{a}T^{a}$, where $T^{a}$ are the $SU(3)$ generators in the fundamental, normalized as
\begin{equation}
\Tr(T^{a}T^{b})=\tfrac12\,\delta^{ab}, \qquad a,b=1,\dots,8.
\label{eq:su3-norm}
\end{equation}
This identity is specific to traceless $3\times 3$ matrices (hence to $SU(3)$) and will be used to reduce quartic adjoint invariants to a single structure.

\subsection*{Cayley--Hamilton proof}
For any traceless $3\times 3$ matrix $A$ the Cayley--Hamilton theorem implies
\begin{equation}
A^{3}-\tfrac12\,\Tr(A^{2})\,A-(\det A)\,\mathds{1}=0.
\label{eq:CH}
\end{equation}
Multiplying \eqref{eq:CH} by $A$ and taking the trace gives
\begin{equation}
\Tr(A^{4})=\tfrac12\big(\Tr A^{2}\big)^{2}.
\label{eq:CH-result}
\end{equation}
Specializing to $A=M$ directly leads to
\begin{equation}
(\Tr M^{2})^{2}=2\,\Tr M^{4},
\end{equation}
which is the desired identity \eqref{eq:main-identity}.

\subsection*{Group--theory proof}
Using the product decomposition
\begin{equation}
T^{a}T^{b}=\tfrac{1}{2N}\,\delta^{ab}\,\mathds{1}
+\tfrac12\!\left(d^{abc}+if^{abc}\right)T^{c},\qquad (N=3),
\label{eq:TT}
\end{equation}
and the fact that $B_{a}B_{b}f^{abc}=0$, one finds
\begin{equation}
M^{2}=\tfrac{B^{2}}{6}\,\mathds{1}+\tfrac12\,S^{c}T^{c},\qquad 
B^{2}\equiv B_{a}B_{a},\quad S^{c}\equiv d^{abc}B_{a}B_{b}.
\label{eq:M2}
\end{equation}
Squaring and tracing gives
\begin{equation}
\Tr M^{4}=\frac{B^{4}}{12}+\frac18\,S^{2},\qquad S^{2}\equiv S^{c}S^{c}.
\label{eq:TrM4-pre}
\end{equation}
For $SU(3)$ the $dd$ identity
\begin{equation}
d^{abe}d^{cde}+d^{ace}d^{bde}+d^{ade}d^{bce}
=\tfrac13\big(\delta^{ab}\delta^{cd}+\delta^{ac}\delta^{bd}+\delta^{ad}\delta^{bc}\big)
\label{eq:dd}
\end{equation}
implies, upon contraction with $B_{a}B_{b}B_{c}B_{d}$, that
\begin{equation}
S^{2}=\tfrac13\,B^{4}.
\label{eq:S2}
\end{equation}
Substituting \eqref{eq:S2} into \eqref{eq:TrM4-pre} yields
\begin{equation}
\Tr M^{4}=\tfrac18\,B^{4}.
\label{eq:TrM4}
\end{equation}
Since $\Tr M^{2}=\tfrac12 B^{2}$, it follows that
\begin{equation}
(\Tr M^{2})^{2}=\tfrac14\,B^{4}=2\,\Tr M^{4},
\label{eq:main-identity-again}
\end{equation}
confirming the result.

\subsection*{Consequence for quartic adjoint invariants}
Because $\Tr M^{4}=\tfrac12(\Tr M^{2})^{2}$ in $SU(3)$, any renormalizable quartic built from one adjoint reduces to a single effective coupling. In particular,
\begin{equation}
g_{1}\,(\Tr M^{2})^{2}+g_{2}\,\Tr(M^{4})
=\Big(g_{1}+\tfrac12\,g_{2}\Big)\,(\Tr M^{2})^{2}.
\label{eq:quartic-coupling}
\end{equation}
Thus there is only one independent four-point coupling constant for adjoint $SU(3)$ fields at dimension four.

\section{Appendix~C: Particle production from a massive vector field}
\label{App-C}

In our model the cubic term in the adjoint potential aligns the vacuum expectation value along
\begin{equation}
v \propto \mathrm{diag}(1,1,-2),
\end{equation}
so that the unbroken subgroup is $U(2)$. The gauge spectrum therefore contains four massive gluons and four massless gluons. The four massive states acquire their longitudinal polarizations by eating the corresponding Goldstone scalars, while the gauge bosons associated with the unbroken $U(2)$ subgroup remain massless.

We work in unitary (Proca) gauge. For the purposes of the present analysis, we retain only the two transverse polarizations of each massive gluon. A proper treatment of the longitudinal modes requires a coupled scalar--vector analysis across the bubble wall and is left for future work.

In Euclidean time $\tilde{\tau}$, each transverse mode satisfies
\begin{equation}
\partial_{\tilde{\tau}}^2 g_i^a(\tilde{\tau},\mathbf{x})
+\nabla^2 g_i^a(\tilde{\tau},\mathbf{x})
-m_g^2(\tilde{\tau})\,g_i^a(\tilde{\tau},\mathbf{x})=0,
\end{equation}
with a sudden mass jump at $\tilde{\tau}=\tilde{\tau}^*$,
\begin{equation}
m_g(\tilde{\tau})=
\begin{cases}
m_g, & \tilde{\tau}<\tilde{\tau}^*,\\
M_g, & \tilde{\tau}>\tilde{\tau}^*.
\end{cases}
\end{equation}
In our case the gluons are massless in the false vacuum and massive in the true vacuum, so
\begin{equation}
m_g=0,
\qquad
M_g\simeq 1~\mathrm{TeV}.
\end{equation}

Decomposing into momentum modes and transverse polarization vectors $\epsilon_i^{(\lambda)}$ gives
\begin{equation}
g_i^a(\tilde{\tau},\mathbf{x})
=
\sum_{\lambda=1}^{2}\int\!\frac{d^3k}{(2\pi)^3}\,
\epsilon_i^{(\lambda)}(\mathbf{k})\,
g_k^{a(\lambda)}(\tilde{\tau})\,
e^{i\mathbf{k}\cdot\mathbf{x}},
\end{equation}
where the sum runs only over the two transverse polarizations. The time-dependent mode function then obeys
\begin{equation}
g_k''(\tilde{\tau})-\bigl[k^2+m_g^2(\tilde{\tau})\bigr]g_k(\tilde{\tau})=0.
\end{equation}
The transverse mode functions are matched in exactly the same way as in the main text for the scalar case, now applied to each transverse polarization of the massive gluon. The general solution is
\begin{equation}
g_k(\tilde{\tau})=
\begin{cases}
e^{\omega_- \tilde{\tau}}, & \tilde{\tau}<\tilde{\tau}^*,\\[6pt]
A_k\,e^{\omega_+ \tilde{\tau}}+B_k\,e^{-\omega_+ \tilde{\tau}}, & \tilde{\tau}>\tilde{\tau}^*,
\end{cases}
\end{equation}
where
\begin{equation}
\omega_-=\sqrt{k^2+m_g^2},
\qquad
\omega_+=\sqrt{k^2+M_g^2}.
\end{equation}

Requiring continuity of both $g_k$ and $\partial_{\tilde{\tau}}g_k$ at $\tilde{\tau}=\tilde{\tau}^*$ yields
\begin{equation}
A_k=
\frac{1}{2\omega_+}\,(\omega_++\omega_-)\,
e^{-(\omega_+-\omega_-)\tilde{\tau}^*},
\qquad
B_k=
\frac{1}{2\omega_+}\,(\omega_+-\omega_-)\,
e^{(\omega_++\omega_-)\tilde{\tau}^*}.
\end{equation}
As in the main text, we set
\begin{equation}
\tilde{\tau}^*=-R_0,
\end{equation}
with $R_0$ the nucleation radius of the bubble.

The particle-production spectrum per mode is then obtained from the corresponding Bogoliubov transform,
\begin{equation}
N_k=
\frac{B_k^2}{A_k^2-B_k^2}
=
\left[
\frac{(\omega_++\omega_-)^2}{(\omega_+-\omega_-)^2}
e^{4\omega_+R_0}
-1
\right]^{-1}.
\end{equation}
This expression applies to each transverse polarization separately. In the phenomenological analysis in the main text, we retain only the two transverse polarizations of each of the four massive gluons. The total transverse gluon contribution is therefore obtained by summing over these eight transverse modes.

\section{Appendix~D: Derivation of
\texorpdfstring{$\Gamma(G\to q\bar q)$}{Gamma(G to q qbar)}}
\label{App-D}

We compute the tree-level partial width for a massive color-octet vector boson
\(G^a_\mu\) (``massive gluon'') of mass \(M_g\) decaying into a Dirac quark of
mass \(m_q\) and its antiquark via the QCD-like interaction. The result is for
one fixed massive adjoint component \(a\). We do not average over adjoint color.

\subsubsection*{Interaction and amplitude}

The interaction Lagrangian is
\begin{equation}
  \mathcal{L}_{\rm int}
  =
  g_s\,\bar q_i\,\gamma^\mu\,T^a_{ij}\,q_j\,G^a_\mu,
\end{equation}
where \(T^a\) are the \(SU(3)\) generators in the fundamental representation,
normalized as
\begin{equation}
\Tr(T^aT^b)=T(R)\delta^{ab},
\qquad
T(R)=\frac12.
\end{equation}
For a fixed adjoint color \(a\), the decay amplitude is
\begin{equation}
  \mathcal{M}
  =
  g_s\,\bar u_i(p)\,\gamma^\mu\,T^a_{ij}\,v_j(p')\,\varepsilon_\mu(k),
  \qquad
  k=p+p',
  \qquad
  k^2=M_g^2.
\end{equation}

\subsubsection*{Spin, polarization, and color algebra}

A massive vector has three physical polarizations. For an unpolarized massive
gluon we average over these three initial polarizations and sum over the final
spins and colors:
\begin{equation}
  \overline{|\mathcal M|^2}
  =
  \frac{1}{3}
  \sum_{\rm spins,\,colors}
  \mathcal{M}\mathcal{M}^\dagger.
\end{equation}
The Proca polarization sum is
\begin{equation}
  \sum_{\lambda=1}^{3}
  \varepsilon_\mu^{(\lambda)}(k)\,
  \varepsilon_\nu^{(\lambda)\,*}(k)
  =
  -g_{\mu\nu}
  +
  \frac{k_\mu k_\nu}{M_g^2}.
  \label{eq:ProcaSum}
\end{equation}
The second term in Eq.~\eqref{eq:ProcaSum} is the longitudinal part of the
massive-vector projector. It does not contribute to the decay amplitude because
the vector current is conserved for equal-mass external spinors. Indeed,
\begin{equation}
  k_\mu \bar u(p)\gamma^\mu v(p')
  =
  \bar u(p)(\slashed p+\slashed p')v(p')
  =
  \bar u(p)(m_q-m_q)v(p')
  =
  0.
  \label{eq:current-conservation-Gqq}
\end{equation}
Therefore the \(k_\mu k_\nu/M_g^2\) term drops out in the squared amplitude,
and the polarization sum may be replaced by \(-g_{\mu\nu}\) after contraction
with the quark current.

This also clarifies the role of the longitudinal polarization. Although the
massive gluon has a longitudinal physical mode, this mode does not generate an
additional Yukawa-like amplitude for \(G\to q\bar q\). This is consistent with
the Goldstone-boson equivalence theorem: in the present model there is no
direct Yukawa coupling between the colored Goldstone modes and the quarks, so
there is no separate Goldstone-mediated enhancement of the longitudinal decay
amplitude. Equivalently, the longitudinal part of the Proca projector gives no
contribution once the vector current is contracted.

The color sum over the final quark and antiquark colors is
\begin{equation}
  \sum_{i,j} T^a_{ij}T^a_{ji}
  =
  \Tr(T^aT^a)
  =
  T(R)
  =
  \frac12,
  \label{eq:color-sum-Gqq}
\end{equation}
where the initial adjoint index \(a\) is kept fixed.

For the spin sums we use
\begin{equation}
  \sum_s u(p)\bar u(p)=\slashed p+m_q,
  \qquad
  \sum_s v(p')\bar v(p')=\slashed p'-m_q.
\end{equation}
The corresponding Dirac trace is
\begin{align}
  \sum_{\rm spins}
  \bar u(p)\gamma^\mu v(p')\,
  \bar v(p')\gamma^\nu u(p)
  &=
  \Tr\!\left[
  (\slashed p+m_q)\gamma^\mu(\slashed p'-m_q)\gamma^\nu
  \right]
  \nonumber\\
  &=
  4\left[
  p^\mu p'^\nu
  +
  p^\nu p'^\mu
  -
  g^{\mu\nu}\left(p\!\cdot\!p'+m_q^2\right)
  \right].
  \label{eq:DiracTrace}
\end{align}
Contracting with \(-g_{\mu\nu}\) gives
\begin{align}
  -g_{\mu\nu}\Tr[\cdots]
  &=
  4\left[
  -p\!\cdot\!p'
  -
  p\!\cdot\!p'
  +
  4\left(p\!\cdot\!p'+m_q^2\right)
  \right]
  \nonumber\\
  &=
  8\left(p\!\cdot\!p'+2m_q^2\right).
  \label{eq:polcontract}
\end{align}
Combining the polarization average, the color factor, and the spin trace, we
obtain
\begin{equation}
  \overline{|\mathcal M|^2}
  =
  \frac{1}{3}\,
  g_s^2\,
  \Tr(T^aT^a)\,
  8\left(p\!\cdot\!p'+2m_q^2\right)
  =
  \frac{4}{3}g_s^2
  \left(p\!\cdot\!p'+2m_q^2\right).
  \label{eq:Mbar2_pp}
\end{equation}

It is useful to express \(p\!\cdot\!p'\) in terms of invariants. Since
\begin{equation}
  (p+p')^2=M_g^2=2m_q^2+2p\!\cdot\!p',
\end{equation}
we have
\begin{equation}
  p\!\cdot\!p'
  =
  \frac12\left(M_g^2-2m_q^2\right).
\end{equation}
Therefore
\begin{equation}
  \overline{|\mathcal M|^2}
  =
  \frac{2}{3}g_s^2
  \left(M_g^2+2m_q^2\right).
  \label{eq:Mbar2_invariants}
\end{equation}

\subsubsection*{Two-body phase space}

For a \(1\to2\) decay into two equal-mass particles, the integrated two-body
phase space is
\begin{equation}
  \int d\Phi_2
  =
  \frac{\beta}{8\pi},
  \qquad
  \beta
  =
  \sqrt{1-\frac{4m_q^2}{M_g^2}}.
\end{equation}
The decay width is therefore
\begin{equation}
  \Gamma
  =
  \frac{1}{2M_g}\,
  \overline{|\mathcal M|^2}\,
  \int d\Phi_2
  =
  \frac{\beta}{16\pi M_g}\,
  \overline{|\mathcal M|^2}.
  \label{eq:GammaMaster}
\end{equation}

\subsubsection*{Final result}

Substituting Eq.~\eqref{eq:Mbar2_invariants} into Eq.~\eqref{eq:GammaMaster}
and using \(g_s^2=4\pi\alpha_s\), we find
\begin{align}
  \Gamma(G\to q\bar q)
  &=
  \frac{\beta}{16\pi M_g}\,
  \frac{2}{3}g_s^2
  \left(M_g^2+2m_q^2\right)
  \nonumber\\[4pt]
  &=
  \frac{4\pi\alpha_s}{24\pi}\,
  M_g
  \left(1+\frac{2m_q^2}{M_g^2}\right)
  \beta
  \nonumber\\[4pt]
  &=
  \frac{\alpha_s}{6}\,
  M_g
  \left(1+\frac{2m_q^2}{M_g^2}\right)
  \sqrt{1-\frac{4m_q^2}{M_g^2}}.
  \label{eq:Gamma-G-qqbar-final}
\end{align}
This formula is for one fixed massive adjoint color component \(a\). It reduces
to
\begin{equation}
\Gamma(G\to q\bar q)=\frac{\alpha_s}{6}M_g
\end{equation}
for massless quarks and vanishes at the threshold \(M_g=2m_q\).

\section{Appendix~E: Derivation of
\texorpdfstring{$\Gamma(G_H\to q\bar q)$}{Gamma(GH to q qbar)}}
\label{App-E}

In this appendix we derive the tree-level partial width for a physical adjoint scalar component $G_H^a$ of mass $M_s$ decaying into a Dirac quark of mass $m_q$ and its antiquark. Since the minimal broken-color model written in the main text does not by itself generate a direct scalar--quark Yukawa interaction, we assume here an effective coupling of the form
\begin{equation}
  \mathcal{L}_{\rm int}
  \;=\;
  -\,y_q\,G_H^a\,\bar q_i\,T^a_{ij}\,q_j,
\end{equation}
where $T^a$ are the $SU(3)$ generators in the fundamental representation, normalized as \cite{Djouadi:2012ua,Gerbush:2007fe}
\begin{equation}
  \Tr(T^a T^b)=T(R)\,\delta^{ab},
  \qquad
  T(R)=\frac12.
\end{equation}
This effective interaction may be viewed as a phenomenological parametrization of the coupling of a given physical scalar mode to quarks after symmetry breaking. In applications one may further choose, for example, a normalization inspired from minimal-flavor-violation $y_q=\eta_q\,m_q/v$, but for the derivation below we keep $y_q$ general.

\subsubsection*{Interaction and amplitude}

For a fixed adjoint color $a$, the decay process
\begin{equation}
  G_H^a(k)\to q_i(p)\,\bar q_j(p'),
  \qquad
  k=p+p',
  \qquad
  k^2=M_s^2,
\end{equation}
has tree-level amplitude
\begin{equation}
  \mathcal{M}
  \;=\;
  y_q\,\bar u_i(p)\,T^a_{ij}\,v_j(p').
\end{equation}
Since the initial particle is a scalar, there is no spin or polarization average.

\subsubsection*{Spin and color algebra}

Summing over the final spins and colors, we obtain
\begin{equation}
  \sum |\mathcal M|^2
  \;=\;
  y_q^2
  \sum_{\rm spins,\,colors}
  \bar u_i(p)\,T^a_{ij}\,v_j(p')\,
  \bar v_{j'}(p')\,T^a_{j'i'}\,u_{i'}(p).
\end{equation}
The color sum is straightforward:
\begin{equation}
  \sum_{i,j}T^a_{ij}T^a_{ji}
  \;=\;
  \Tr(T^aT^a)
  \;=\;
  T(R)
  \;=\;
  \frac12.
\end{equation}
For the spin sums we use
\begin{equation}
  \sum_s u(p)\bar u(p)=\slashed p+m_q,
  \qquad
  \sum_s v(p')\bar v(p')=\slashed p'-m_q,
\end{equation}
which gives
\begin{align}
  \sum_{\rm spins} \bar u(p)v(p')\,\bar v(p')u(p)
  &=
  \Tr\!\left[(\slashed p+m_q)(\slashed p'-m_q)\right]
  \nonumber\\
  &=
  4\,(p\!\cdot\!p'-m_q^2).
\end{align}
Combining the spin and color factors, we get
\begin{equation}
  \sum |\mathcal M|^2
  \;=\;
  y_q^2 \,\Tr(T^aT^a)\,4\,(p\!\cdot\!p'-m_q^2)
  \;=\;
  2\,y_q^2\,(p\!\cdot\!p'-m_q^2).
\end{equation}
It is useful to rewrite this in terms of invariants. Since
\begin{equation}
  (p+p')^2=M_s^2=2m_q^2+2\,p\!\cdot\!p',
\end{equation}
we have
\begin{equation}
  p\!\cdot\!p'=\frac{M_s^2-2m_q^2}{2},
\end{equation}
and therefore
\begin{equation}
  p\!\cdot\!p'-m_q^2
  \;=\;
  \frac{M_s^2-4m_q^2}{2}.
\end{equation}
The squared matrix element then becomes
\begin{equation}
  \sum |\mathcal M|^2
  \;=\;
  y_q^2\,(M_s^2-4m_q^2).
  \label{eq:GHqq_M2}
\end{equation}

\subsubsection*{Two-body phase space}

For a scalar decaying into two equal-mass fermions, the integrated two-body phase space is
\begin{equation}
  \int d\Phi_2
  \;=\;
  \frac{\beta_q}{8\pi},
  \qquad
  \beta_q\equiv\sqrt{1-\frac{4m_q^2}{M_s^2}}.
\end{equation}
The standard decay-width formula is therefore
\begin{equation}
  \Gamma
  \;=\;
  \frac{1}{2M_s}\,\sum |\mathcal M|^2\,\int d\Phi_2
  \;=\;
  \frac{\beta_q}{16\pi M_s}\,\sum |\mathcal M|^2.
\end{equation}
Substituting \eqref{eq:GHqq_M2}, we obtain
\begin{equation}
  \Gamma(G_H^a\to q\bar q)
  \;=\;
  \frac{\beta_q}{16\pi M_s}\,y_q^2\,(M_s^2-4m_q^2).
\end{equation}
Using
\begin{equation}
  M_s^2-4m_q^2=M_s^2\,\beta_q^2,
\end{equation}
the result simplifies to
\begin{equation}
  \Gamma(G_H^a\to q\bar q)
  \;=\;
  \frac{y_q^2\,M_s}{16\pi}\,\beta_q^3.
\end{equation}

\subsubsection*{Final result}

The tree-level partial width for a fixed adjoint scalar component $G_H^a$ decaying to a quark--antiquark pair is therefore
\begin{equation}
  \Gamma(G_H^a\to q\bar q)
  \;=\;
  \frac{y_q^2\,M_s}{16\pi}
  \left(1-\frac{4m_q^2}{M_s^2}\right)^{3/2}.
\end{equation}
If one adopts the phenomenological normalization
\begin{equation}
  y_q=\eta_q\,\frac{m_q}{v},
\end{equation}
with $v=246~{\rm GeV}$, the width may be written as
\begin{equation}
  \Gamma(G_H^a\to q\bar q)
  \;=\;
  \frac{\eta_q^2\,m_q^2\,M_s}{16\pi v^2}
  \left(1-\frac{4m_q^2}{M_s^2}\right)^{3/2}.
\end{equation}
This formula is for a fixed adjoint index $a$ and vanishes at threshold $M_s=2m_q$.

\section{Appendix~F: Derivation of
\texorpdfstring{$\Gamma(G_H\to gg)$}{Gamma(GH to gg)}}
\label{App-F}

In this appendix we derive the one-loop partial width for a physical adjoint scalar component $G_H^a$ of mass $M_s$ decaying into two gluons through quark triangle diagrams. We again assume the effective Yukawa interaction
\begin{equation}
  \mathcal{L}_{\rm int}
  \;=\;
  -\,y_q\,G_H^a\,\bar q_i\,T^a_{ij}\,q_j,
\end{equation}
together with the standard QCD coupling
\begin{equation}
  \mathcal{L}_{\rm QCD}
  \;=\;
  g_s\,\bar q_i\gamma^\mu T^b_{ij}q_j\,g^b_\mu.
\end{equation}
The decay $G_H^a\to g^b g^c$ then proceeds through the usual quark triangle, summed over the quark flavors circulating in the loop \cite{Gresham:2007ri, Logan:2014jla}.

\subsubsection*{Color structure of the triangle amplitude}

For a fixed incoming scalar color $a$ and outgoing gluon colors $b,c$, the two triangle orderings produce the color factor
\begin{equation}
  \Tr(T^aT^bT^c)+\Tr(T^aT^cT^b)
  \;=\;
  \Tr\!\bigl(T^a\{T^b,T^c\}\bigr).
\end{equation}
Using
\begin{equation}
  \{T^b,T^c\}
  \;=\;
  \frac{1}{3}\delta^{bc}\,\mathds{1}
  + d^{bcd}T^d,
\end{equation}
together with $\Tr(T^a)=0$, one obtains
\begin{equation}
  \Tr\!\bigl(T^a\{T^b,T^c\}\bigr)
  \;=\;
  d^{bcd}\,\Tr(T^aT^d)
  \;=\;
  \frac12\,d^{abc}.
\end{equation}
Thus the decay amplitude is proportional to the totally symmetric $SU(3)$ tensor $d^{abc}$, as expected for a CP-even scalar coupled to two gauge bosons.

\subsubsection*{General form of the amplitude}

Gauge invariance fixes the Lorentz structure of the on-shell amplitude to be transverse with respect to both external gluon momenta. It is therefore convenient to write the amplitude as
\begin{equation}
  i\mathcal{M}^{abc}_{\mu\nu}
  \;=\;
  i\,d^{abc}\,C_s\,
  \Big[(p_1\!\cdot\!p_2)\,g_{\mu\nu}-p_{2\mu}p_{1\nu}\Big],
  \label{eq:GHgg_amp_general}
\end{equation}
where $p_1^2=p_2^2=0$ and $(p_1+p_2)^2=M_s^2$. The loop coefficient $C_s$ contains the quark masses, Yukawa couplings, and the standard scalar triangle form factor. Summing over all quark flavors in the loop, one finds
\begin{equation}
  C_s
  \;=\;
  \frac{\alpha_s}{6\pi}\,
  \sum_q \frac{y_q}{m_q}\,F_{1/2}(\tau_q),
  \qquad
  \tau_q\equiv \frac{4m_q^2}{M_s^2}.
\end{equation}
Here $F_{1/2}(\tau)$ is the usual spin-$\tfrac12$ scalar loop form factor. A convenient representation, using
\(\tau_q=4m_q^2/M_{s}^2\), is
\begin{equation}
  F_{1/2}(\tau)
  =
  \frac{3}{2}\,\tau
  \left[
  1+(1-\tau)f(\tau)
  \right],
\end{equation}
with
\begin{equation}
  f(\tau)=
  \begin{cases}
    \arcsin^2\!\bigl(1/\sqrt{\tau}\bigr), & \tau\ge 1,\\[1ex]
    -\dfrac14
    \left[
      \ln\!\left(\dfrac{1+\sqrt{1-\tau}}{1-\sqrt{1-\tau}}\right)-i\pi
    \right]^2, & \tau<1.
  \end{cases}
\end{equation}
This normalization is such that \(F_{1/2}(\tau)\to1\) in the heavy-fermion
limit \(\tau\to\infty\).

\subsubsection*{Polarization and color sums}

To obtain the squared amplitude we sum over the two transverse polarizations of each final-state gluon and over the final-state colors $b,c$. Defining
\begin{equation}
  T_{\mu\nu}
  \;\equiv\;
  (p_1\!\cdot\!p_2)\,g_{\mu\nu}-p_{2\mu}p_{1\nu},
\end{equation}
the squared matrix element becomes
\begin{equation}
  \sum |\mathcal M|^2
  \;=\;
  \sum_{b,c,\rm pol}
  d^{abc}d^{abc}\,|C_s|^2\,T_{\mu\nu}T^{\mu\nu}.
\end{equation}
For on-shell massless gluons,
\begin{equation}
  p_1^2=p_2^2=0,
  \qquad
  p_1\!\cdot\!p_2=\frac{M_s^2}{2},
\end{equation}
and the polarization sum yields
\begin{equation}
  \sum_{\rm pol} T_{\mu\nu}T^{\mu\nu}
  \;=\;
  2\,(p_1\!\cdot\!p_2)^2
  \;=\;
  \frac{M_s^4}{2}.
\end{equation}
For a fixed initial adjoint index $a$, the color identity
\begin{equation}
  \sum_{b,c} d^{abc}d^{abc}
  \;=\;
  \frac53
\end{equation}
then gives
\begin{equation}
  \sum |\mathcal M|^2
  \;=\;
  \frac53\,|C_s|^2\,\frac{M_s^4}{2}.
\end{equation}

\subsubsection*{Two-body phase space}

The final state consists of two identical massless bosons. The integrated two-body phase space therefore gives
\begin{equation}
  \Gamma
  \;=\;
  \frac{1}{32\pi M_s}\,\sum |\mathcal M|^2.
\end{equation}
Substituting the expression above, one obtains
\begin{equation}
  \Gamma(G_H^a\to gg)
  \;=\;
  \frac{1}{32\pi M_s}\,
  \frac53\,|C_s|^2\,\frac{M_s^4}{2}
  \;=\;
  \frac{5M_s^3}{192\pi}\,|C_s|^2.
\end{equation}
Inserting
\begin{equation}
  C_s
  \;=\;
  \frac{\alpha_s}{6\pi}\,
  \sum_q \frac{y_q}{m_q}\,F_{1/2}(\tau_q),
\end{equation}
we arrive at the final result
\begin{equation}
  \Gamma(G_H^a\to gg)
  \;=\;
  \frac{5\,\alpha_s^2\,M_s^3}{6912\,\pi^3}
  \left|
    \sum_q \frac{y_q}{m_q}\,F_{1/2}(\tau_q)
  \right|^2
  .
\end{equation}

\subsubsection*{Phenomenological normalization}

If the effective Yukawa coupling is written as
\begin{equation}
  y_q=\eta_q\,\frac{m_q}{v},
  \qquad
  v=246~{\rm GeV},
\end{equation}
then the width takes the form
\begin{equation}
  \Gamma(G_H^a\to gg)
  \;=\;
  \frac{5\,\alpha_s^2\,M_s^3}{6912\,\pi^3\,v^2}
  \left|
    \sum_q \eta_q\,F_{1/2}(\tau_q)
  \right|^2.
\end{equation}
For the benchmark mass used in the main text, $M_s=2.5~{\rm TeV}$, the dominant contribution generally comes from the top-quark loop if the coefficients $\eta_q$ are flavor-hierarchical in the usual Yukawa sense. Nevertheless, because the process is loop-induced, the $gg$ channel is naturally suppressed compared to the tree-level decay into quark--antiquark pairs whenever the latter is kinematically open.

\section{\texorpdfstring{{Appendix~G: Justification of the cutoffs and finite-temperature consistency of the production estimates}}{Appendix G: Justification of the cutoffs and finite-temperature consistency of the production estimates}}
\label{App-G}

In this appendix we justify the different stopping prescriptions and consistency
conditions used for the vacuum--mismatch, subcritical thermal, and wall--matter
transition-radiation channels in the \(SU(3)_c\to U(2)\) benchmark. The three
mechanisms are controlled by different physical quantities. The
vacuum--mismatch source is controlled by the wall proper acceleration
\(\alpha(\tau)\). The thermal source is controlled by the finite-temperature
phase diagram, the shocked-layer temperature, and the freeze-out temperature
\(T_*<T_c\). The transition-radiation source is controlled by the flux of
ambient matter through the wall and by the wall-frame energy of the incident
quarks.

\subsection*{Vacuum--mismatch production: exponential cutoff}

The vacuum--mismatch channel is governed by the zero-mode occupancy,
Eq.~\eqref{eq:Nk0-alpha-su3c},
\begin{equation}
N_{k=0}(\tau)
=
\left[
\frac{(\omega_+ + \omega_-)^2}{(\omega_+ - \omega_-)^2}
\exp\!\Bigl(\frac{4\omega_+}{\alpha(\tau)}\Bigr)-1
\right]^{-1}.
\end{equation}
For small positive \(\alpha(\tau)\), this reduces to
\begin{equation}
N_{k=0}(\tau)
\simeq
\left[
\frac{(\omega_+ + \omega_-)^2}{(\omega_+ - \omega_-)^2}
\right]^{-1}
\exp\!\left(-\frac{4\omega_+}{\alpha(\tau)}\right).
\end{equation}
Thus the source is exponentially suppressed once the acceleration falls below
the particle mass scale. In the late-time regime we may approximate
\begin{equation}
\alpha(\tau)\simeq \alpha(0)\,e^{-\tau/\tau_{\rm term}},
\end{equation}
so that at \(\tau=5\tau_{\rm term}\),
\begin{equation}
\alpha(5\tau_{\rm term})
=
\alpha(0)e^{-5}
\simeq
1.07\times10^{-3}~{\rm TeV}.
\end{equation}
For the scalar channel, \(\omega_+=M_s=2.5~{\rm TeV}\), giving
\begin{equation}
\frac{4M_s}{\alpha(5\tau_{\rm term})}
\simeq
9.29\times10^3,
\qquad
N_{k=0}^{(s)}(5\tau_{\rm term})
\sim e^{-9.29\times10^3}\ll 10^{-4000}.
\end{equation}
For the massive-gluon channel, \(\omega_+=M_g=1.0~{\rm TeV}\), giving
\begin{equation}
\frac{4M_g}{\alpha(5\tau_{\rm term})}
\simeq
3.75\times10^3,
\qquad
N_{k=0}^{(g)}(5\tau_{\rm term})
\sim e^{-3.75\times10^3}\ll 10^{-1600}.
\end{equation}
Hence by \(5\tau_{\rm term}\) the vacuum--mismatch source is completely
extinguished in both channels. Even though the area factor grows polynomially,
this polynomial growth cannot compete with the exponential suppression. This
justifies terminating the vacuum--mismatch channel once the wall reaches the
terminal regime.

This is also why the swept-volume reinterpretation used below for the thermal
channel is not separately applied to vacuum mismatch. In the vacuum-mismatch
calculation, the production history is already integrated over the expanding
wall through
\begin{equation}
\frac{dN_{\rm tot}}{d\tau}
=
\mathcal{N}_0\,
N_{k=0}(\tau)\,
4\pi R^2(\tau)\sinh y(\tau).
\end{equation}
where $\mathcal{N}_0$ is the normalization constant. The swept volume enters through \(4\pi R^2\,dR\), but the local source is not a
thermal density. It is the acceleration-dependent occupation \(N_{k=0}\), which
rapidly shuts off as \(\alpha(\tau)\) decreases.

\subsection*{Finite-temperature cutoff}

The thermal channel cannot be extended to arbitrarily high temperature in the
broken phase. The finite-temperature effective potential gives
\begin{equation}
T_c\simeq0.256~{\rm TeV},
\qquad
T_{\rm sp}\simeq0.416~{\rm TeV},
\end{equation}
while the massive broken-phase particles have
\begin{equation}
M_g\simeq1.0~{\rm TeV},
\qquad
M_s\simeq2.48~{\rm TeV}.
\end{equation}
Thus
\begin{equation}
T_c<M_g<M_s.
\end{equation}
An equilibrated thermal bath hot enough to make the massive gluons or scalar
octets ultrarelativistic would therefore lie in the symmetry-restored regime
and should not be described as a broken-phase gas of massive modes.

\subsection*{Transition radiation: terminal motion and environment dependence}

The transition-radiation channel is not controlled by the wall acceleration and
does not require a thermal bath above \(T_c\). It is controlled by the flux of
ambient matter crossing the wall and by the wall-frame energy of the incident
quarks,
\begin{equation}
E_q'\simeq x_q\gamma_w m_N.
\end{equation}
Therefore this channel can continue even after the wall reaches terminal
motion. Its total yield scales as
\begin{equation}
N^{\rm rad}\propto n_bD^3,
\end{equation}
where \(n_b\) is the ambient baryon density and \(D\) is the distance over which
the bubble sweeps matter. The unknown efficiency factors \(f_G\) and \(f_H\)
encode finite wall thickness, color projection, hadronic coherence, and
possible multiple-emission effects. Thus the transition-radiation yield should
be viewed as an order-of-magnitude environment-dependent estimate, not as a
universal microscopic prediction.

\section{Appendix~H: Proper Acceleration of the Bubble Wall}\label{App-H}

We derive here the magnitude of the proper acceleration of the bubble wall described by the equation \( r^2 - t^2 = R_0^2 \).

The bubble wall follows a hyperbolic trajectory in \((t, r)\) spacetime, which can be parametrized using proper time \(\tau\) (the time experienced by an observer on the bubble wall). The proper parametrization is:
\begin{equation}
t = R_0 \sinh\left(\frac{\tau}{R_0}\right), \quad r = R_0 \cosh\left(\frac{\tau}{R_0}\right).
\end{equation}
This satisfies the equation \( r^2 - t^2 = R_0^2 \), as:
\begin{equation}
R_0^2 \cosh^2\left(\frac{\tau}{R_0}\right) - R_0^2 \sinh^2\left(\frac{\tau}{R_0}\right) = R_0^2.
\end{equation}

The 4-velocity \( u^\mu \) of the bubble wall is the derivative of the coordinates with respect to proper time
\begin{equation}
u^\mu = \left( \frac{dt}{d\tau}, \frac{dr}{d\tau} \right) = \left( \cosh\left(\frac{\tau}{R_0}\right), \sinh\left(\frac{\tau}{R_0}\right) \right).
\end{equation}
Thus, the norm of the 4-velocity is
\begin{equation}
u^\mu u_\mu = \left(\frac{dt}{d\tau}\right)^2 - \left(\frac{dr}{d\tau}\right)^2 = \cosh^2\left(\frac{\tau}{R_0}\right) - \sinh^2\left(\frac{\tau}{R_0}\right) = 1,
\end{equation}
confirming that \(\tau\) is the proper time.

The 4-acceleration \( a^\mu \) is the derivative of the 4-velocity with respect to proper time
\begin{equation}
a^\mu = \frac{du^\mu}{d\tau} = \left( \frac{1}{R_0} \sinh\left(\frac{\tau}{R_0}\right), \frac{1}{R_0} \cosh\left(\frac{\tau}{R_0}\right) \right).
\end{equation}
The norm of the 4-acceleration (which gives the proper acceleration) is
\begin{equation}
a^\mu a_\mu = \left(\frac{1}{R_0} \sinh\left(\frac{\tau}{R_0}\right)\right)^2 - \left(\frac{1}{R_0} \cosh\left(\frac{\tau}{R_0}\right)\right)^2 = -\frac{1}{R_0^2}.
\end{equation}
The magnitude of the proper acceleration is the absolute value of the norm \cite{Walker:1984vj,Good:2013lca}
\begin{equation}
|a| = \sqrt{|a^\mu a_\mu|} = \frac{1}{R_0}.
\end{equation}

Thus, in the absence of friction, the magnitude of the proper acceleration of the bubble wall is constant and given by
\begin{equation}
\dfrac{1}{R_0}.
\end{equation}

\section{Appendix~I: Numerical procedure}\label{App-I}
We evolve the coupled system $\{y(\tau),t(\tau),R(\tau),N_{\rm tot}(\tau)\}$ defined by \eqref{Parameters-su3c}–\eqref{eq:dNtot} using a fourth–order Runge–Kutta (RK4) method with adaptive steps. The state is initialized at nucleation by
\begin{equation}
y(0)=0\;(v=0),\qquad t(0)=0,\qquad R(0)=R_0,\qquad N_{\rm tot}(0)=0.
\label{eq:init-cond}
\end{equation}
At each substep we compute the instantaneous acceleration
\begin{equation}
\alpha_{\rm raw}(\tau)=A-\frac{2}{R(\tau)}-B\sinh y(\tau).
\label{eq:alpha-raw-SU3}
\end{equation}
If $\alpha_{\rm raw}\le0$, we enforce terminal balance by freezing $dy/d\tau=0$ and setting $N_{k=0}=0$; otherwise the system evolves as
\begin{equation}
\frac{dy}{d\tau}=\alpha_{\rm raw}, \ \ \ 
\frac{dt}{d\tau}=\cosh y, \ \ \ 
\frac{dR}{d\tau}=\sinh y, 
\end{equation}
with particle production
\begin{equation}
\frac{dN_{\rm tot}}{d\tau}
=
\mathcal{N}_0\,
N_{k=0}(\tau)\,
4\pi R(\tau)^2\sinh y(\tau),
\label{eq:dNtot-SU3}
\end{equation}
Here \(\mathcal{N}_0\) is an effective zero-mode phase-space normalization with dimensions of \({\rm TeV}^3\). In the numerical estimates below we set \(\mathcal{N}_0=1~{\rm TeV}^3\).
and the instantaneous occupancy
\begin{equation}
N_{k=0}(\tau)
=\left[\frac{(\omega_++\omega_-)^2}{(\omega_+-\omega_-)^2}\,
e^{\tfrac{4\omega_+}{\alpha_{\rm raw}}}-1\right]^{-1}.
\label{eq:Nk0-SU3}
\end{equation}

\noindent\textit{Stability.} 
For large arguments, specifically when
\[
\frac{4\omega_+}{\alpha_{\rm raw}}\gtrsim200,
\]
the stable asymptotic approximation is used:
\begin{equation}
N_{k=0}(\tau)
\simeq
\left[\frac{(\omega_++\omega_-)^2}{(\omega_+-\omega_-)^2}
\right]^{-1}e^{-4\omega_+/\alpha_{\rm raw}}.
\end{equation}
The proper-time step is adapted according to the local drive,
\begin{equation}
d\tau \;\sim\; \frac{0.05}{\max(|\alpha_{\rm raw}|,10^{-6})},
\label{eq:dtau-adapt-SU3}
\end{equation}
and clamped to $20\le d\tau\le 800$, ensuring stability both in the curvature-dominated early phase and in the slow terminal approach.

\noindent\textit{Stopping and checks.}  
Integration halts when
$
\tau=\tau_{\rm term}.
\label{eq:stop-cond-SU3}
$
Convergence is checked by halving the initial $d\tau$ and demanding relative differences $<10^{-4}$ in $\{\tau_{\rm term},R_{\rm fin},N_{\rm tot}^{\rm (int)}\}$. Consistency checks include verifying the exact acceleration evolution,
\begin{equation}
\frac{d\alpha}{d\tau}
=
\frac{2\sinh y}{R^2}
-
B\cosh y\,\alpha,
\end{equation}
the monotonic growth of \(R(\tau)\), the approach of \(\alpha(\tau)\) toward zero in the terminal regime, and \(N_{k=0}=0\) whenever \(\alpha_{\rm raw}\le0\).
 To illustrate this, we study ultra–relativistic terminal deficits $\delta=1-v_{\rm term}$, each mapped to a friction pair $(\eta,\tau_{\rm term})$ fixing $B=\eta/\sigma$. The self–consistent balance is enforced by
\begin{equation}
\gamma_{\rm term}\,v_{\rm term}=\frac{\Delta V}{\eta}, 
\label{eq:balance-SU3}
\end{equation}
so that for each $\delta$,
\begin{equation}
\eta(\delta)=\frac{\Delta V}{\gamma_{\rm term}v_{\rm term}}, 
\qquad
\tau_{\rm term}=\frac{\sigma}{\eta}.
\label{eq:eta-tau-SU3}
\end{equation}
This ensures the chosen terminal velocity matches the proper-time dynamics of the SU(3)\(_c\) bubble wall.

\bibliographystyle{JHEP}
\bibliography{References}

\end{document}